\documentclass[aoas,preprint]{imsart}

\RequirePackage[OT1]{fontenc}
\RequirePackage[numbers]{natbib}

\usepackage{amsthm,amsmath}
\usepackage{amsbsy}

\input xy
\xyoption{all}

\usepackage{algorithm}
\usepackage[noend]{algpseudocode}

\usepackage{epsf}
\usepackage{epsfig}

\startlocaldefs
\newcommand{\cov}{\mbox{Cov}}
\newcommand{\bfU}{\boldsymbol{U}}
\newcommand{\bfW}{\boldsymbol{W}}
\newcommand{\bfL}{\boldsymbol{L}}
\newcommand{\bfH}{\boldsymbol{H}}

\newcommand{\ind}{1\!\!1}
\newcommand{\ci}{\perp\!\!\!\perp}
\newcommand{\nci}{\not\!\perp\!\!\!\perp}
\newcommand{\bfA}{\boldsymbol{A}}
\newcommand{\bfB}{\boldsymbol{B}}
\newcommand{\bfC}{\boldsymbol{C}}

\newcommand{\bfa}{\boldsymbol{a}}
\newcommand{\bfb}{\boldsymbol{b}}

\newcommand{\bfV}{\boldsymbol{V}}

\newcommand{\mG}{\mathcal{G}}
\endlocaldefs

\newcommand{\beginsupplement}{%
        \setcounter{table}{0}
        \renewcommand{\thetable}{S\arabic{table}}%
        \setcounter{figure}{0}
        \renewcommand{\thefigure}{S\arabic{figure}}%
     }

\begin{document}

\begin{frontmatter}

\title{Towards personalized causal inference of medication response in mobile health: an instrumental variable approach for randomized trials with imperfect compliance}
\thankstext{}{$^1$Sage Bionetworks. $^2$Fred Hutchinson Cancer Research Center and University of Washington, Seattle}
\runtitle{Personalized causal inference in mobile health}


\author{\fnms{Elias} \snm{Chaibub Neto$^1$, Ross L Prentice$^2$, Brian M Bot$^1$, Mike Kellen$^1$, Stephen H Friend$^1$, Andrew D Trister$^1$, Larsson Omberg$^1$, Lara Mangravite$^1$}\ead[label=e1]{elias.chaibub.neto@sagebase.org, Sage Bionetworks}}
\address{\printead{e1}}


\runauthor{Chaibub Neto E. et. al.}

\begin{abstract}
Mobile health studies can leverage longitudinal sensor data from smartphones to guide the application of personalized medical interventions. In this paper, we propose that adoption of an instrumental variable approach for randomized trials with imperfect compliance provides a natural framework for personalized causal inference of medication response in mobile health studies. Randomized treatment suggestions can be easily delivered to the study participants via electronic messages popping up on the smart-phone screen. Under quite general assumptions we can identify the causal effect of the actual treatment on the response in the presence of unobserved confounders. We implement a personalized randomization test of the null hypothesis of no causal effect of treatment on response, and evaluate its performance in a large scale simulation study encompassing data generated from linear and non-linear time series models under several simulation conditions. In particular, we evaluate the empirical power of the proposed test under varying degrees of compliance between the suggested and actual treatment adopted by the participant. Our investigations provide encouraging results in terms of power and control of type I error rates. Finally, we compare the proposed instrumental variable approach to a simple intent-to-treat strategy, and develop randomization confidence intervals for the causal effects.
\end{abstract}

\end{frontmatter}

\section{Introduction}

Mobile health platforms are becoming a popular tool for the implementation of precision medicine programs. The goal is to leverage longitudinal sensor data, collected by smart-phones or activity tracking devices, to better inform the application of personalized medical interventions. In particular, this approach can allow the evaluation of treatment efficacy for an individual participant, as opposed to the traditional modus operandi of medicine, where the efficacy of a treatment is evaluated in a clinical trial performed over an specific cohort of patients and, hence, can only establish treatment efficacy at a population level (Topol 2012, Schork 2015).

\begin{figure}[!h]
\begin{center}
\includegraphics[angle=270, scale = 0.6, clip]{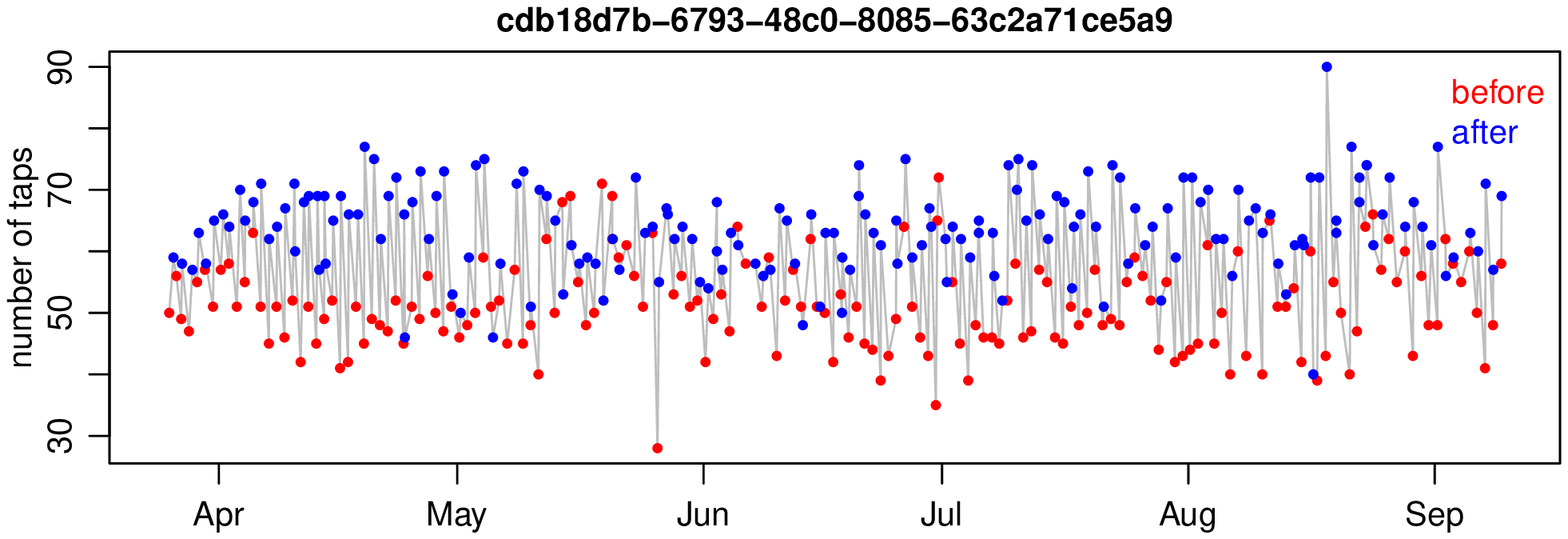}
\caption{Example of data collected by the mPower study. The figure shows the data from a single study participant, collected over a six months period, and color coded according to whether the number of taps was recorded when the participant was medicated (blue dots) or not (red dots).}
\label{supplefig:numberTaps}
\end{center}
\end{figure}

The personalized medicine application motivating the present work comes from the mPower study (Trister \textit{et al} 2016, Bot \textit{et al} 2016), one of the studies launched with Apple's ResearchKit mobile platform (Friend 2015). In this purely observational study, a participant is asked to perform activity tasks, including tapping, voice, memory, posture and gait tests. Raw sensor data collected from each test is processed into activity specific features, which represent objective measures of the current state of the patient's disease. For instance, the number of times a participant can tap the screen of a smartphone over a period of 20 seconds represents one such feature, where lower number of taps indicate a more severe state. Since the activity tests are performed by the patient on a daily basis, before and after medication, over several months, the processed data corresponds to time series of feature measurements annotated according to whether the measurement was taken before or after the patient has taken medication. Supplementary Figure \ref{supplefig:numberTaps} shows an example.

This personalized medicine problem is clinically relevant since the determination of whether or not a Parkinson patient is responding to its current medication can help the physician make more informed treatment recommendations for the patient. But since mPower is an observational study, it is challenging to draw any causal conclusions about medication effect due to the potential of unobserved confounders.

The inference of causal effects at the personalized level is especially vulnerable to cyclical confounding effects, defined as any recurrent patterns or fluctuations in the response variable that are not caused by the treatment (Beasley \textit{et al} 1997). The standard remedy to deal with such confounding issues is to randomize the treatment schedule, which in the personalized context (where a single participant is followed over time) boils down to randomly assigning the treatments over time. In other words, the experimental units correspond to the same study participant at different points in time. However, it would be naive to expect a study participant to faithfully follow an assigned treatment schedule. In order to address this problem, in this paper we propose and evaluate the statistical properties of an instrumental variable approach (Angrist and Krueger 2001, Greenland 2000, Didelez \textit{et al} 2010, Baiocchi \textit{et al} 2014) for longitudinally randomized trials with imperfect compliance.

In the context of our motivating problem, an instrumental variable (IV) corresponds to a randomized treatment suggestion, prompting the participant to perform the activity task either before or after taking the medication. Such, randomized treatment suggestions can be easily delivered to the study participants via electronic messages displayed on the smart-phone screen. Section 3 presents our proposed IV approach, with detailed descriptions of the assumptions required for the identification of the causal effect in our motivating application.

In order to test the null hypothesis of no causal effect between treatment and response, we propose a randomization test (Section 3.1) and evaluate its performance in a large scale simulation study (Section 4). We also, point out that while the randomization test based on the IV estimator statistic is exactly equivalent to a randomization test based on a simple intention-to-treat statistic (Fisher \textit{et al} 1990) (Section 5), it turns out that, in the context of our motivating mobile health application, the estimates from the intent-to-treat analysis tend to be biased towards zero, so that the causal effect estimates generated by the IV approach might outperform the intention-to-treat estimates, in data sets where the treatment effect is different from zero. This observation suggests that, in practice, the IV approach might be more appealing, especially in face of the current trend in the biomedical field where researchers are encouraged to report parameter estimates and confidence intervals, in addition to p-values. To meet this need, we also develop randomization confidence intervals for the causal effects by inverting randomization tests (Section 6). Finally, in Section 7 we discuss our results.

The next section presents general definitions, notation, and background material on causal inference and instrumental variables.

\section{Background}

\subsection{General definitions and notation}

Throughout this paper, we consider longitudinal data indexed by $t = 1, \ldots, n$, where $Z_t$ represents a binary instrumental variable assuming the value 1 if the electronic suggestion asks the participant to perform the activity task after taking medication, and 0 if it asks the participant to perform the activity task before medication (the treatment assignment mechanism corresponds to a Bernoulli trial with probability of success equal to $P(Z_t = 1) = 0.5$); $X_t$ is a binary treatment variable set to 1 if the participant actually performs the activity task after taking medication, and to 0 if the participant performs the activity task before taking medication (i.e., $X_t = 1$ if the participant is medicated, and $X_t = 0$ if he/she is unmedicated); and $Y_t$ represents a real valued response variable representing an extracted feature from the raw activity task data (e.g., number of taps in a fixed time interval). We represent the set of observed confounders of $X_t$ and $Y_t$ by $\bfW_t$. We denote by $\bfU_t$ the set of time specific unmeasured confounders affecting both $X_t$ and $Y_t$. The set of ubiquitous latent variables, which influence the $Y_t$ measurements across all time points, is denoted by $\bfL$. Similarly, we let $\bfH$ represent the set of ubiquitous latent variables influencing the $X_t$ measurements. Finally, we denote the set of ubiquitous confounders of $X_t$ and $Y_t$ by $\bfC$. We adopt a direct acyclic graph (DAG) representation of the dynamic causal process underlying the observed and unobserved variables. We reserve the symbols E(), Var(), Cov() and Cor() for the expectation, variance, covariance and correlation operators, respectively. Statistical independence and dependence are represented, respectively, by the symbols $\ci$ and $\nci$, while conditional independence relations are described by the notation $\bfA_1 \ci \bfA_2 \mid \bfA_3$ meaning that the set of variables $\bfA_1$ is independent of the set $\bfA_2$ conditional on the set $\bfA_3$. The set difference between sets $\bfA_1$ and $\bfA_2$ is expressed as $\bfA_1 \setminus \bfA_2$. We let $\ind\{ A \}$ represent the indicator function assuming value 1 if event $A$ occurs, and 0 otherwise.

\subsection{Stationary time series}

In time series analysis, the concept of stationarity captures the notion of regularity over time in the probabilistic behavior of the series (Shumway and Stoffer 2011). A strictly stationary time series is defined as one for which the probabilistic behavior of every collection of variables, $\{ Y_{1}, Y_{2}, \ldots, Y_{k} \}$, is identical to the shifted collection, $\{ Y_{1 + j}, Y_{2 + j}, \ldots, Y_{k + j} \}$, for all $k = 1, 2, \ldots$, all time points $1, \ldots, k$, and all shifts $j = 0, \pm 1, \pm 2, \ldots$. A milder version of stationarity (more often assumed in practice) only imposes restrictions in the first two moments of the series, that is, the mean value of the series is constant and independent of $t$, and the autocovariance is a function of the shift $j$ and not of time directly.

The stationarity assumption plays a critical role in the analysis of time series data, since we do not typically have an independent and identically distributed sample, $\{ Y_{t,1}$, $Y_{t,2}$, $\ldots$, $Y_{t,n_t}\}$, of the variable $Y_{t}$, but rather a single observation at each data point $Y_{t}$. In this situation, with a single realization per time point, the assumption of stationarity allows us to compute standard sample statistics using the time series data (Shumway and Stoffer 2011). For instance, we can compute the mean value of the time series using $n^{-1} \sum_{t=1}^{n} y_t$.

\subsection{Causal inference}

Following Pearl (2000), we adopt a mechanism-based account of causation. In this framework, the statistical information about a set of variables, encoded by the joint probability distribution, is supplemented by a causal DAG encoding a qualitative description of our assumptions about the causal relation between the variables. The joint probability distribution factorizes according to the causal DAG structure,
\begin{equation}
P\big(x_1, \ldots, x_p \big) = \prod_{j} P\big(x_j \mid pa(x_j)\big)~,
\label{eq:factorizationformula}
\end{equation}
where each element, $P\big(x_j \mid pa(x_j)\big)$, represents an autonomous mechanism describing the relationship between variable $X_j$ and its parents. A non-parametric representation of these elements is given by $x_j = h_j(pa(x_j), \epsilon_j)$, where $h_j$ represents a deterministic function of the parents of $X_j$ and a random disturbance term $\epsilon_j$. In this framework, causation means predicting the consequences of an intervention over a set of variables in the DAG, where intervention is expressed as a ``surgery" on the equations and associated causal graph.

We use the $do$ operator notation to distinguish $P\big(y \mid do(X = x)\big)$ from  $P(y \mid X = x)$, where the former quantity describes the post-intervention distribution of a variable $Y$ given that the value of $X$ was set be $x$ by an external intervention, while the latter represents the usual conditional distribution of $Y$ given that we observed the value of $X$ to be equal to $x$ (and is denoted the observational or pre-intervention distribution). For interventions over a single variable, the relationship between the pre-intervention and post-intervention distributions is given by the truncated factorization formula,
\begin{equation}
P\big(x_1, \ldots, x_p \mid do(X_k = x_k')\big) = \prod_{j \not= k} P\big(x_j \mid pa(x_j)\big) \, \ind\{ x_k = x_k' \}~,
\label{eq:factorizationformula}
\end{equation}
where the removal of the equation $P\big(x_k \mid pa(x_k)\big)$ from the product in equation (\ref{eq:factorizationformula}), and the replacement of $x_k$ by $x_k'$ in all elements $P\big(x_j \mid pa(x_j)\big)$ for which $X_k$ is a parent of $X_j$, formalizes what is meant by an ``intervention surgery".

The causal effect of intervention $X$ on the $Y$ is usually defined as a function of the post-intervention distribution $P\big(y \mid do(X = x)\big)$. In this paper we adopt the average causal effect of $X$ on $Y$ defined as,
\begin{equation}
\mbox{ACE}(X \rightarrow Y) = \mbox{E}\big( Y \mid do(X = x_2) \big) - \mbox{E}\big( Y \mid do(X = x_1) \big)~,
\end{equation}
where $x_1$ is usually some baseline value. We say that a causal effect of $X$ on $Y$ is identifiable if the post-intervention distribution $P\big(y \mid do(X = x)\big)$, and hence the $\mbox{ACE}(X \rightarrow Y)$ quantity, is a function of observed variables only.

\subsection{Instrumental variables}

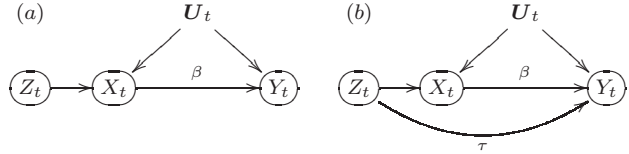
\begin{figure}[!h]
{\scriptsize
$$
\xymatrix@-0.75pc{
(a) && \bfU_t \ar[dl] \ar[dr] && (b) && \bfU_t \ar[dl] \ar[dr] & \\
*+[F-:<10pt>]{Z_t} \ar[r] & *+[F-:<10pt>]{X_t} \ar[rr]^{\beta} && *+[F-:<10pt>]{Y_t} & *+[F-:<10pt>]{Z_t} \ar[r] \ar@/_1.5pc/[rrr]_{\tau} & *+[F-:<10pt>]{X_t} \ar[rr]^{\beta} && *+[F-:<10pt>]{Y_t} \\
}
$$}
\caption{In panel a, $Z_t$ can potentially qualify as an instrumental variable for the identification of the causal effect $\beta$, since: (\textit{i}) $Z_t$ is marginally independent of $\bfU_t$, as readily seem by application of the d-separation criterion to the DAG structure; (\textit{ii}) there is an arrow from $Z_t$ to $X_t$; and (\textit{iii}) there is no direct arrow from $Z_t$ to $Y_t$, and the the indirect causal effect of $Z_t$ on $Y_t$ is mediated exclusively by $X_t$. In panel b $Z_t$ does not qualify as an instrument for the identification of $\beta$ because assumption \textit{iii} is violated.}
\label{fig:simpleexample}
\end{figure}

When it is not possible to rule out the existence of unmeasured confounders affecting both treatment and response variables, it is still possible to use an instrumental variable to identify the causal effect of the treatment on the response, whenever certain parametric and distributional assumptions hold. The DAG in Figure \ref{fig:simpleexample}a provides a graphical representation of three necessary (although not sufficient) assumptions (Didelez \textit{et al} 2010) for the identification of the causal effect $\beta$, namely: (\textit{i}) $Z_t$ is marginally independent of all unmeasured confounders which influence both treatment and response variables, that is $Z_t \ci \bfU_t$; (\textit{ii}) $Z_t$ must be statistically associated with $X_t$, that is $Z_t \nci X_t$;
and (\textit{iii}) any association between $Z_t$ and $Y_t$ must be exclusively mediated by $X_t$, that is, conditionally on $X_t$ and $\bfU_t$, $Z_t$ and $Y_t$ must be independent, $Z_t \ci Y_t \mid \{X_t, \bfU_t\}$.

\section{Instrumental variables for longitudinal randomized trials with imperfect compliance in the context of mobile health}

In observational studies, assumptions \textit{i} to \textit{iii} need to be carefully evaluated in order to assess the validity of the putative instrument. However, in the context of randomized clinical trials with imperfect compliance in mobile health, assumption \textit{i} is valid by construction due to the randomization of the assigned suggestions, which effectively makes variable $Z_t$ statistically independent of any measured or unmeasured confounder of the treatment/outcome relation at time $t$. Assumption \textit{ii} is valid if there is some degree of compliance between the randomly assigned treatment suggestions and the treatment effectively adopted by the study participants. (In practice, assumption \textit{ii} is likely to hold since the treatment suggestion does not seem to increase the amount of burden to the study participant.) Assumption \textit{iii}, also known as the exclusion restriction, is only guaranteed to hold in double-blinded trials (Hernan and Robins 2006). In the context of our motivating application, this assumption is not guaranteed to hold, since a reminder might change the participant behavior in other ways that affect the outcome other than through the treatment (for instance, condition \textit{iii} would be violated if the receipt of a reminder prompted a participant to take a co-medication other than the one under study). In any case, assumption \textit{iii} seems to be at least approximately reasonable in the proposed application. (Although in the particular case that a participant takes medication only once per day and around the same time every day, it is possible that this condition might be violated, as described in Appendix A. Nevertheless, Appendix A also describes a strategy for minimizing this potential source of bias.)

In the following we show that the identification of the causal effect (from observed data) holds quite generally in the context of linear and non-linear time series models, under the additional assumptions that $X_t$ is linearly associated with $Y_t$, the causal effect $\beta$ is constant over time, and that the $Y_t$ and $X_t$ time series are stationary. Note that we only require a linear association between $Y_t$ and $X_t$, without making any assumptions about the relationships between $Y_t$ and all other measured covariates, unmeasured confounders, and lagged response variables, or about the serial dependency structure over the $X_t$ measurements, or about the relationship between $Z_t$ and $X_t$ and between $Z_t$ and $Y_t$.

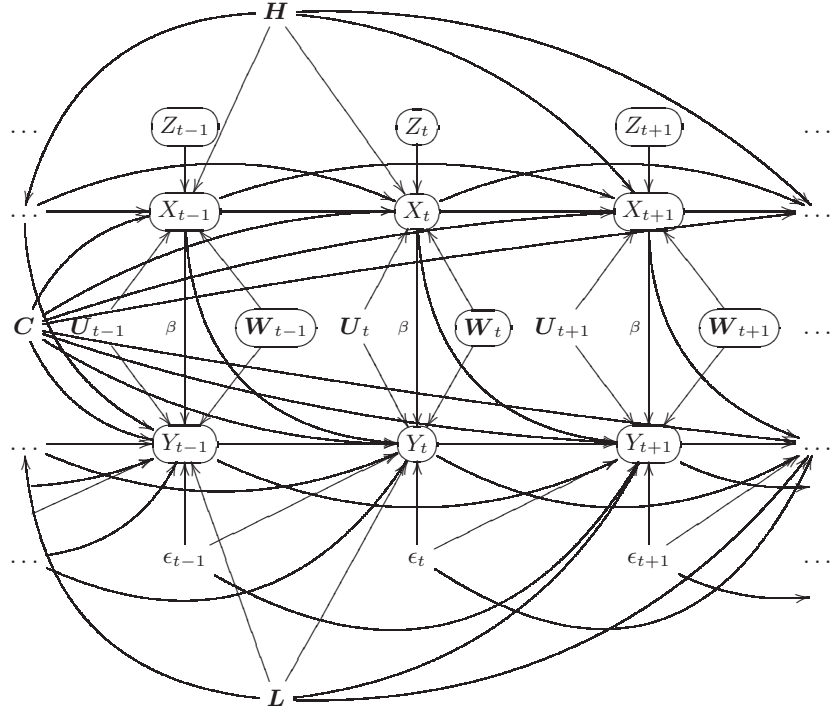
\begin{figure}[!h]
{\footnotesize
$$
\xymatrix@-1.5pc{
&&& \bfH \ar@/_2.0pc/[dddddlll] \ar[dddddl] \ar[dddddrr] \ar@/^1.5pc/[dddddrrrrr] \ar@/^2.0pc/[dddddrrrrrrr] &&&&&&&& \\
&&&&&&&&&&& \\
&&&&&&&&&&& \\
\ldots & & *+[F-:<10pt>]{Z_{t-1}} \ar[dd] & & & *+[F-:<10pt>]{Z_t} \ar[dd] & & & *+[F-:<10pt>]{Z_{t+1}} \ar[dd] && \ldots \\
&&&&&&&&&&& \\
\ldots \ar@/^1.5pc/[rrrrr] \ar@/_1.5pc/[ddddddrr] \ar[rr] & & *+[F-:<10pt>]{X_{t-1}} \ar[dddddd]_\beta \ar[rrr] \ar@/_2.5pc/[ddddddrrr] \ar@/^1.5pc/[rrrrrr] & & & *+[F-:<10pt>]{X_t} \ar[dddddd]_\beta \ar[rrr] \ar@/_2.5pc/[ddddddrrr] \ar@/^1.5pc/[rrrrr] & & & *+[F-:<10pt>]{X_{t+1}} \ar[dddddd]_\beta \ar[rr] \ar@/_1.75pc/[ddddddrr] && \ldots \\
&&&&&&&&&&& \\
&&&&&&&&&&& \\
\bfC \ar@/^1pc/[uuurr] \ar@/_1pc/[dddrr] \ar@/^1.0pc/[uuurrrrr] \ar@/_1.0pc/[dddrrrrr] \ar@/^0.75pc/[uuurrrrrrrr] \ar@/_0.75pc/[dddrrrrrrrr] \ar@/^0.25pc/[uuurrrrrrrrrr] \ar@/_0.25pc/[dddrrrrrrrrrr] & \bfU_{t-1} \ar[uuur] \ar[dddr] & & *+[F-:<10pt>]{\bfW_{t-1}} \ar[uuul] \ar[dddl] & \bfU_t \ar[uuur] \ar[dddr] & & *+[F-:<10pt>]{\bfW_t} \ar[uuul] \ar[dddl] & \bfU_{t+1} \ar[uuur] \ar[dddr] & & *+[F-:<10pt>]{\bfW_{t+1}} \ar[uuul] \ar[dddl] & \ldots \\
&&&&&&&&&&& \\
&&&&&&&&&&& \\
\ldots \ar[rr] \ar@/_1.5pc/[rrrrr] & & *+[F-:<10pt>]{Y_{t-1}} \ar[rrr] \ar@/_2pc/[rrrrrr] & & & *+[F-:<10pt>]{Y_t} \ar[rrr] \ar@/_2pc/[rrrrr] & & & *+[F-:<10pt>]{Y_{t+1}} \ar[rr] \ar@/_0.5pc/[drr] && \ldots \\
\ar@/_0.3pc/[urr] &&&&&&&&&&& \\
\ar[uurr] &&&&&&&&&&& \\
\ldots \ar@/_1pc/[uuurr] \ar@/_3pc/[uuurrrrr] & & \epsilon_{t-1} \ar[uuu] \ar[uuurrr] \ar@/_4pc/[uuurrrrrr] & & & \epsilon_t \ar[uuu] \ar[uuurrr] \ar@/_4pc/[uuurrrrr] & & & \epsilon_{t+1} \ar[uuu] \ar[uuurr] \ar@/_0.5pc/[drr] && \ldots \\
&&&&&&&&&&& \\
&&&&&&&&&&& \\
&&&&&&&&&&& \\
&&& \bfL \ar@/^2.5pc/[uuuuuuulll] \ar[uuuuuuul] \ar[uuuuuuurr] \ar@/_2pc/[uuuuuuurrrrr] \ar@/_2.5pc/[uuuuuuurrrrrrr] &&&&&&&& \\
}
$$}
\caption{DAG representation of a complex dynamic model. The response variable, $Y_t$, follows an autoregressive and moving average process of order 2 (i.e., is influenced by $Y_{t-1}$ and $Y_{t-2}$, on the autoregressive component, and by  $\epsilon_{t-1}$ and $\epsilon_{t-2}$ in the moving average part), and is further influenced by observed covariates ($\bfW_t$, $X_t$, $X_{t-1}$) and unobserved variables ($\bfU_t$, $\bfL$, $\bfC$). The treatment variable, $X_t$, follows an autoregressive process of order 2 (i.e., is influenced by $X_{t-1}$ and $X_{t-2}$, and is also influence by observed covariates ($\bfW_t$), and unobserved variables ($\bfU_t$, $\bfH$, $\bfC$), in addition to the instrument, $Z_t$.}
\label{fig:generalarma}
\end{figure}

To fix ideas, consider the complex dynamic model presented in Figure \ref{fig:generalarma}, which will be used as a concrete example in the following argument. Under the assumption that $Y_t$ and $X_t$ are linearly associated and $\beta$ is constant over time, a general time series model is given by,
\begin{equation}
Y_t = \beta \, X_t + f(pa(Y_t) \setminus X_t)~,
\label{eq:generalts}
\end{equation}
where the $pa(Y_t)$ represents the set of parents of variable $Y_t$, and $f()$ represents a general function of the variables in $pa(Y_t) \setminus X_t$. In principle, the variables in $pa(Y_t) \setminus X_t$ might include: time specific observed covariates and unobserved confounders up to time point $t$ (e.g., $\bfW_{t}, \bfW_{t-1}, \ldots$, and $\bfU_{t}, \bfU_{t-1}, \ldots$); unobserved variables (e.g., $\bfL$ and $\bfC$); lagged treatment and response variables up to time point $t-1$ (e.g., $X_{t-1}, X_{t-2}, \ldots$ and $Y_{t-1}, Y_{t-2},$ $\ldots$); lagged error terms up to time point $t$ (e.g., $\epsilon_{t}, \epsilon_{t-1}, \ldots$). For the particular example in Figure \ref{fig:generalarma} we have that $pa(Y_t) \setminus X_t = \{ \bfW_t, \bfU_t, \bfL, \bfC,$ $X_{t-1}, Y_{t-1}, Y_{t-2}, \epsilon_t, \epsilon_{t-1}$, $\epsilon_{t-2} \}$.

Because in our mobile health application the instrumental variable $Z_t$ is randomized, we have, by construction, that $Z_t$ is independent of all variables in the set $pa(Y_t) \setminus X_t$, and hence independent of any function $f()$ of the variables in $pa(Y_t) \setminus X_t$. Therefore, it follows that,
\begin{align}
\cov(Z_t, Y_t) &= \beta \, \cov(Z_t, X_t) + \cov(Z_t, f(pa(Y_t) \setminus X_t)) \\ \nonumber
&= \beta \, \cov(Z_t, X_t)~,
\end{align}
so that we can identify the causal effect,
\begin{equation}
\beta \, = \, \frac{\cov(Z_t, Y_t)}{\cov(Z_t, X_t)}~, \hspace{0.3cm} \mbox{for all $t = 1, \ldots, n$,}
\label{eq:causaleffect}
\end{equation}
from observed data.

Hence, for any time series model which can be represented by equation (\ref{eq:generalts}) we have that, for any time point $t$, the causal effect of $X_t$ on $Y_t$ can be estimated as the ratio of the covariance estimates. The problem, however, is that we have a single measurement for $Y_t$, $X_t$ and $Z_t$ per time point $t$, and not a sample $\{ (Z_{t,1}, X_{t,1}, Y_{t,1}), \ldots,$ $(Z_{t,n_t}, X_{t,n_t}, Y_{t,n_t}) \}$ of measurements of $(Z_t, X_t, Y_t)$. Furthermore, both $X_t$ and $Y_t$ time series might show serial dependencies. Nonetheless, if the time series over the $Y_t$ and $X_t$ variables are stationary (so that the statistical properties of $Y_t$, $X_t$ and $Z_t$ variables are similar across all $t$ indexes), then we can estimate the (constant) causal effect in equation (\ref{eq:causaleffect}) using the data from all time points via the standard sample covariance estimator,
$$
\hat{\beta}_{IV} = \frac{\widehat{\cov}(Z_t, Y_t)}{\widehat{\cov}(Z_t, X_t)}
$$
\begin{equation}
= \frac{n^{-1} \sum_{t=1}^{n} Z_{t} Y_{t} - (n^{-1} \sum_{t=1}^{n} Z_{t}) (n^{-1} \sum_{t=1}^{n} Y_{t})}{n^{-1} \sum_{t=1}^{n} Z_{t} X_{t} - (n^{-1} \sum_{t=1}^{n} Z_{t}) (n^{-1} \sum_{t=1}^{n} X_{t})}~.
\label{eq:causaleffectestimator}
\end{equation}

At this point one might indicate that the above estimator (\ref{eq:causaleffectestimator}) is only valid under the assumptions that $Z_t$ is linearly associated with $X_t$ and $Y_t$, since the covariance operator only captures linear associations between two variables, and it is possible that two variables have zero covariance when the first variable has a causal influence on a second one mediated by a non-linear mechanism (so that the covariance operator fails to capture the non-linear association pattern between the variables). We point out, however, that this potential issue cannot happen in our motivating application since both $Z_t$ and $X_t$ are binary variables, and it can be shown (see Appendix B) that an estimate of the non-parametric average causal effect of $Z_t$ on $Y_t$,
$$
\widehat{\mbox{ACE}}(Z_t \rightarrow Y_t) = \widehat{E}\big(Y_t \mid do(Z_t = 1)\big) - \widehat{E}\big(Y_t \mid do(Z_t = 0)\big)
$$
\vspace{-0.5cm}
\begin{equation}
= \frac{n^{-1} \sum_{t=1}^{n} Z_{t} Y_{t} - (n^{-1} \sum_{t=1}^{n} Z_{t}) (n^{-1} \sum_{t=1}^{n} Y_{t})}{(n^{-1} \sum_{t=1}^{n} Z_{t}) (1 - n^{-1} \sum_{t=1}^{n} Z_{t})}
\label{eq:aceZonY}
\end{equation}
is proportional to $\widehat{\cov}(Z_t, Y_t)$, and that an estimate of the non-parametric causal effect of $Z_t$ on $X_t$,
$$
\widehat{\mbox{ACE}}(Z_t \rightarrow X_t) = \widehat{E}\big(X_t \mid do(Z_t = 1)\big) - \widehat{E}\big(X_t \mid do(Z_t = 0)\big) \label{eq:aceZonX}
$$
\vspace{-0.5cm}
\begin{equation}
= \frac{n^{-1} \sum_{t=1}^{n} Z_{t} X_{t} - (n^{-1} \sum_{t=1}^{n} Z_{t}) (n^{-1} \sum_{t=1}^{n} X_{t})}{(n^{-1} \sum_{t=1}^{n} Z_{t}) (1 - n^{-1} \sum_{t=1}^{n} Z_{t})}
\label{eq:aceZonX}
\end{equation}
is proportional to $\widehat{\cov}(Z_t, X_t)$, and that the estimator in (\ref{eq:causaleffectestimator}) actually corresponds to the ratio of the non-parametric causal effects in (\ref{eq:aceZonY}) and (\ref{eq:aceZonX}), showing that the estimator in (\ref{eq:causaleffectestimator}) is still valid without the linearity assumptions (although its validity still requires additivity in the errors and unmeasured confounders).

In addition to confounding, selection bias is another major obstacle to the validity of causal inference in clinical studies. Appendix C presents a discussion of selection bias in the context of personalized mobile health and shows examples where selection mechanisms can lead to bias in the medication effect estimates (as well as, situations where selection does not bias the IV estimates).

\subsection{Randomization test}

We implemented a randomization test (Ernst 2004) for testing the sharp null hypothesis, $H_0: \beta = 0$, against the alternative $H_1: \beta \not= 0$. The randomization null distribution is generated by evaluating the statistic $\hat{\beta}_{IV}$ in (\ref{eq:causaleffectestimator}) on a large number of shuffled versions of the data, where the $Y_t$ measurements are shuffled relative to the $(Z_t, X_t)$ measurements (whose connection is kept intact in order to preserve the association between the $Z_t$ and $X_t$ variables).

\section{Simulation study}

In order to evaluate the statistical properties of the proposed randomization test, we performed a large scale simulation study comprised of 80 separate simulation experiments involving 10 distinct linear and non-linear time series models described in Table 1, and 8 distinct simulation settings described in Table 2.

\begin{table}[!h]
\caption{\label{tab:tsmodels}Time series models used in the simulation study.}
\centering
\begin{tabular}{ll}
\hline
name & response model \\ \hline
ARMA(1, 1) & $Y_t = g + \phi_1 \, Y_{t-1} + \theta_1 \, \epsilon_{t - 1} +  \epsilon_{t}$ \\
ARMA(1, 0) & $Y_t = g + \phi_1 \, Y_{t-1} +  \epsilon_{t}$ \\
ARMA(0, 1) & $Y_t = g + \theta_1 \, \epsilon_{t - 1} +  \epsilon_{t}$ \\
ARMA(0, 0) & $Y_t = g +  \epsilon_{t}$ \\
ARCH(1) & $Y_t = g + \epsilon_{t} \, \sigma_t~, \hspace{0.3cm} \sigma_t^2 = \mu_{\sigma} + a_1 \, Y_{t-1}^2$ \\
GARCH(1, 1) & $Y_t = g + \epsilon_{t}  \, \sigma_t~, \hspace{0.3cm} \sigma_t^2 = \mu_{\sigma} + a_1 \, Y_{t-1}^2 + b_1 \sigma_{t-1}^2$ \\
TAR(1) & $Y_t = g + \phi_{1,1} \, Y_{t-1} \ind\{T_t \leq 0\} + \phi_{1,2} \, Y_{t-1} \ind\{T_t > 0\} + \epsilon_{t}$ \\
LSTAR(1) & $Y_t = g + \phi_{1,1} \, Y_{t-1} G(T_t) + \phi_{1,2} \, Y_{t-1} (1 - G(T_t)) + \epsilon_{t}$ \\
& $G(t_t) = 1/(1 + e^{-t_t})$ \\
ESTAR(1) & $Y_t = g + \phi_{1,1} \, Y_{t-1} G(T_t) + \phi_{1,2} \, Y_{t-1} (1 - G(T_t)) + \epsilon_{t}$ \\
& $G(t_t) = 1 - e^{-t_t^2}$ \\
SETAR(1) & $Y_t = g + \phi_{1,1} \, Y_{t-1} \ind\{Y_{t-1} \leq 0\} + \phi_{1,2} \, Y_{t-1} \ind\{Y_{t-1} > 0\} + \epsilon_{t}$ \\
& \\
& where $g = \lambda \, W_t + \eta \, U_t + \psi \, L + \beta \, X_t + \delta_1 \, X_{t-1}$ \\
\hline
\end{tabular}
\end{table}

\begin{table}[!h]
\caption{\label{tab:simsettings}Distinct settings used in the simulation study.}
\centering
\begin{tabular}{llcc}
\hline
setting & error type & data simulated under & dependency for $X_t$ \\ \hline
1: & $N(0, 1)$ & $H_1: \beta \not= 0$ & complex \\
2: & $U(-\sqrt{3}, \sqrt{3})$ & $H_1: \beta \not= 0$ & complex \\
3: & $N(0, 1)$ & $H_0: \beta = 0$ & complex \\
4: & $U(-\sqrt{3}, \sqrt{3})$ & $H_0: \beta = 0$ & complex \\
5: & $N(0, 1)$ & $H_1: \beta \not= 0$ & simple \\
6: & $U(-\sqrt{3}, \sqrt{3})$ & $H_1: \beta \not= 0$ & simple \\
7: & $N(0, 1)$ & $H_0: \beta = 0$ & simple \\
8: & $U(-\sqrt{3}, \sqrt{3})$ & $H_0: \beta = 0$ & simple \\
\hline
\end{tabular}
\end{table}

The time series models included: autoregressive moving average (ARMA) models (Box \textit{et al} 1994); autoregressive conditional heteroskedasticity (ARCH) models (Engle 1982); generalized autoregressive conditional heteroskedasticity (GARCH) models (Bollerslev 1986); threshold autoregressive (TAR) models (Tong 1978); self exciting threshold autoregressive (SETAR) models (Tong and Lim 1980); and logistic and exponential smooth transition autoregressive (LSTAR and STAR) models (Van Dijk 2002). The 8 distinct simulation settings in Table 2 comprise all possible combinations of simulations generated: under the null or alternative hypothesis; using gaussian or uniform error terms in the generation of the continuous variables, $\epsilon_t$, $U_t$, $W_t$, $L$, and $H$; and adopting either a complex or a simple dependency structure for the $X_t$ variables. For the complex dependency structure, the $X_t$ variables were simulated according to,
\begin{equation}
X_t = \ind\{ \alpha \, Z_t + \omega \, W_t + \gamma \, U_t + \varphi \, H + \varepsilon_{t}^\ast > 0 \}~,
\label{eq:thresholdmodelXtdependent}
\end{equation}
with the error term $\varepsilon_{t-1}^\ast$ generated according to the AR(1) process,
\begin{equation}
\varepsilon_{t}^\ast = \rho \, \varepsilon_{t-1}^\ast + \varepsilon_{t}~, \hspace{0.3cm}  \varepsilon_{t} \sim \mbox{N}(0, 1) \hspace{0.2cm} \mbox{or} \hspace{0.2cm} \varepsilon_{t} \sim \mbox{U}(-\sqrt{3}, \sqrt{3})~,
\end{equation}
whereas for the simple dependency structure,
\begin{equation}
X_t = \ind\{ \alpha \, Z_t + \omega \, W_t + \gamma \, U_t + \varepsilon_{t} > 0 \}~.
\label{eq:thresholdmodelXtindependent}
\end{equation}
Note that under the complex dependency structure the $X_t$ measurements are dependent due to the effect of the common latent variable $H$ and to the serial association induced by the AR(1) process underlying the $\varepsilon_{t}^\ast$ error terms. Under the simple dependency structure, on the other hand, the $X_t$ measurements are independent. Observe, as well, that in the simulations employing uniform distributions, we adopted the range $[-\sqrt{3}, \sqrt{3}]$ in order to ensure that the variance is still 1. For the TAR, LSTAR, and ESTAR models, we generated the threshold variable, $T_t$, from a standard normal distribution.

Each one of the 80 distinct simulation experiments were based on 10,000 simulated data sets. Each simulated data set was generated using a unique combination of simulation parameter values. Table 3 presents the ranges of the simulation parameter values employed in the study. We selected a wide range, $[-4, 4]$, for model parameters $\beta$, $\omega$, $\gamma$, $\varphi$, $\lambda$, $\eta$, $\psi$, $\delta_1$, $\theta_1$. The range of $\alpha$ was strictly positive since this parameter controls the amount of compliance between $Z_t$ and $X_t$, which is assumed to be positive. The range for parameters $\phi_1, \phi_{1,1}, \phi_{1,2}, \rho$ was set to $[-0.8, 0.8]$, since these parameters control autoregressive processes of order 1, and need to be constrained between $[-1, 1]$ in order to ensure the stationarity of the time series. The parameter $a_1$ controls the autoregressive processes over the variance terms in a ARCH(1) or GARCH(1, 1) process and was allowed to vary between $[0, 0.99]$ in our simulations. For GARCH(1, 1) models the additional moving average parameter $b_1$ is set to $1 - a_1$ in order to ensure that the stationarity condition $a_1 + b_1 < 1$ holds. The range of sample size parameter, $n$, was set to realistic values we expect to see in practice. In order to select parameter values spread as uniformly as possible over the entire parameter range we employed a Latin hypercube design (Santner \text{et al} 2003), optimized according to the maximin distance criterium (Johnson \textit{et al} 1990), in the determination of the parameter values used on each of the 10,000 simulated data sets for each of the 80 simulation experiments. In total, our simulations encompassed 800,000 simulated data sets.

\begin{table}[!h]
\caption{\label{tab:simparranges}Simulation parameter ranges.}
\centering
\begin{tabular}{lccccc}
\hline
parameter &&&&& range \\ \hline
$\alpha$ &&&&& $[0.5 \, , \, 4]$ \\
$\beta \, , \, \omega \, , \, \gamma \, , \, \varphi \, , \, \lambda \, , \, \eta \, , \, \psi \, , \, \delta_1 \, , \, \theta_1$ &&&&& $[-4 \, , \, 4]$ \\
$\phi_1 \, , \, \phi_{1,1} \, , \, \phi_{1,2} \, , \, \rho$ &&&&& $[-0.8 \, , \, 0.8]$ \\
$a_1$ &&&&& $[0 \, , \, 0.99]$ \\
$n$ &&&&& $\{ 50, 51, 52, \ldots, 800 \}$ \\
\hline
\end{tabular}
\end{table}

In order to evaluate if adjustment for observed confounders would improve the performance of the randomization test, we compared straight causal effect estimates against adjusted estimates, where the $X_t$ and $Y_t$ variables are replaced by the residuals of the regressions of these variables on the observed confounder $W_t$. Also, in order to illustrate the importance of employing the instrumental variable approach in the presence of unobserved confounders, we compare the straight and adjusted IV approaches against naive tests for the null $H_0: \beta = 0$, based on standard and adjusted t-tests (where, again, we replace the $X_t$ and $Y_t$ by residuals in the adjusted t-test).

\begin{figure}[!h]
\begin{center}
\includegraphics[angle=270, scale = 0.63, clip]{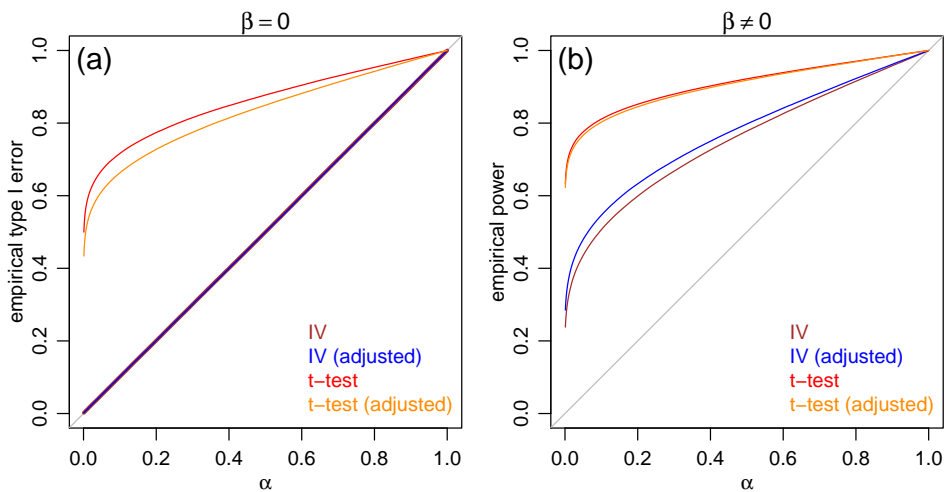}
\caption{Comparison of empirical type I error rates and empirical power.}
\label{fig:pulledresults}
\end{center}
\end{figure}

Supplementary Figure S2a presents the distributions of the autocorrelations between $Y_{t-1}$ and $Y_t$, for all 10 models in Table 1. Supplementary Figure S2b reports the distributions of the correlations between the instrumental variable and all other variables. Figure \ref{fig:pulledresults}a presents the empirical type I error rates, as a function of the nominal level $\alpha$, for all 400,000 data sets simulated under $H_0: \beta = 0$. The plot clearly shows that the randomization test for the straight and adjusted IV approaches (brown and blue curves) is able to control the type I error rates at the nominal levels. Use of t-tests, on the other hand, lead to highly inflated error rates, since these naive approaches mistakenly detect the presence of a causal effect whenever the treatment and response variables are associated entirely because of the influence of observed and unobserved confounders. The plot also shows that the adjustment for the observed covariate (orange curve) is able to reduce the type I error rate by accounting for part of the association between treatment and response variables, as illustrated by the drop in error rate from the red (un-adjusted) to the orange (adjusted) t-tests.

Figure \ref{fig:pulledresults}b presents the empirical power, as a function of $\alpha$, for all 400,000 data sets simulated under $H_1: \beta \not= 0$. As expected, the dominance of the blue curve over the brown one shows that adjustment for observed confounders can improve the power of the randomization test to detect a causal effect. The plot also shows that the t-tests are better powered than the randomization test to detect a causal effect when one exists. We point out, however, that this increased power is an artifact of the biased estimates for $\beta$ delivered by the naive approaches, as clearly illustrated in Supplementary Figure S3, where the $\hat{\beta}$ estimates generated by the naive approaches tend to show larger bias than the estimates generated by the instrumental variable approach.

In order to evaluate the power of the IV approach under varying amounts of compliance by the study participants, and under different sample sizes and strengths of the causal effects, we present in Figure \ref{fig:powerbycorxzweak} (and in Supplementary Figures S4 and S5) empirical power curves stratified according to the correlation between the instrumental and treatment variables, for 4 disjoint sample size intervals, $50 \leq n < 200$, $200 \leq n < 400$, $400 \leq n < 600$, and $600 \leq n \leq 800$, when the causal effect is weak (i.e., $|\beta| < 1$, Figure \ref{fig:powerbycorxzweak}), moderate (i.e., $1 \leq |\beta| < 3$, Supplementary Figure S4), and strong (i.e., $3 \leq |\beta| \leq 4$, Supplementary Figure S5). Inspection of the three plots shows, as one would expect, that for a fixed compliance level the power increases with increasing sample sizes and increasing strength (in absolute value) of the causal effects. Additionally, for any given panel the power increases as a function of the amount of compliance.

\begin{figure}[!h]
\begin{center}
\includegraphics[angle=270, scale = 0.63, clip]{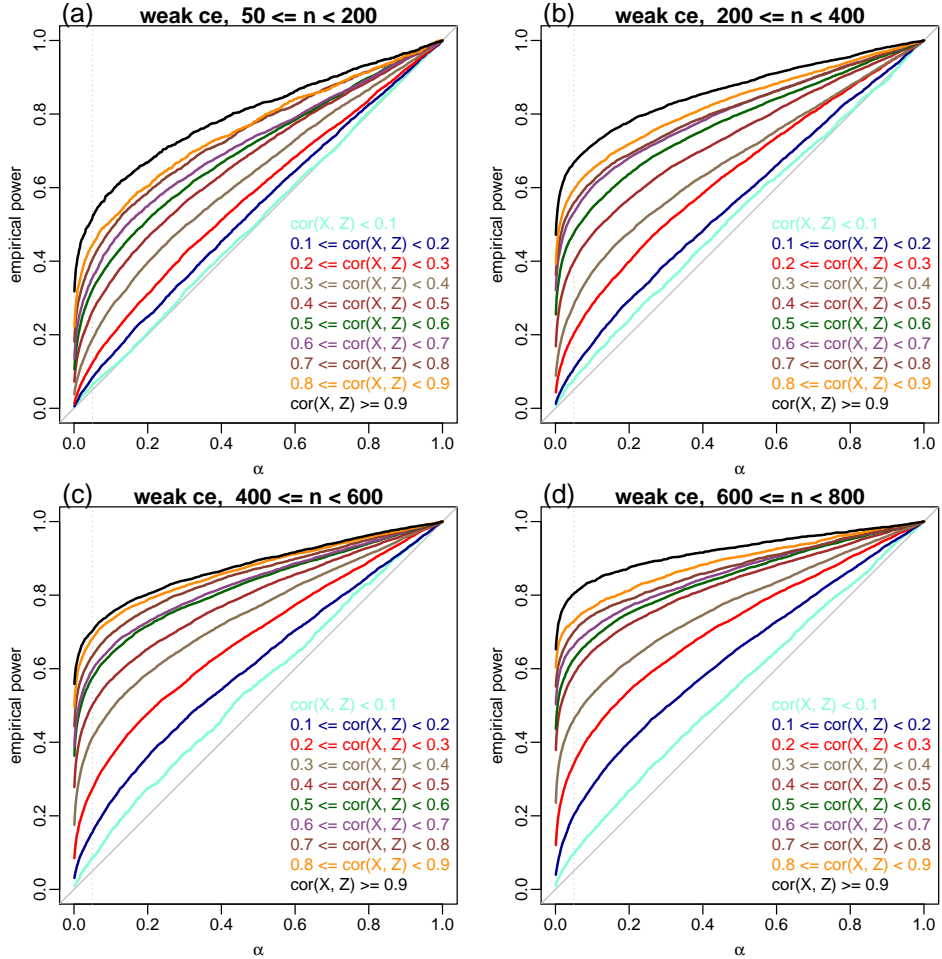}
\caption{Empirical power curves for the adjusted IV causal effect estimator stratified according to cor(X, Z) and to the sample size, n, for simulated data sets generated with weak causal effect, $|\beta| < 1$. The vertical dotted line is set at $\alpha = 0.05$.}
\label{fig:powerbycorxzweak}
\end{center}
\end{figure}

It is important to point out that weak compliance levels not only lead to under-powered tests, but can also lead to highly biased estimates of causal effect when the estimated covariance between instrumental and treatment variables is close to zero (see Supplementary Figure S6 for an illustration). Hence, in practice, it is necessary to check the IV assumption that the instrumental and treatment variables are statistically associated if the main goal is to estimate the causal effect. We point out, however, that violations of this assumption do not not lead to inflation of type I error rates, as illustrated in Supplementary Figure S7b. Finally, we point out that violation of stationarity does not lead to inflated type I error rates, but can wipe out the power to detect a causal effect, as illustrated in Supplementary Figure S8. The R code (R Core Team 2014) implementing the randomization tests and confidence intervals, and used in the generation of the simulation results and figures is available at: https://www.synapse.org/mhealthIV.

\section{Comparison with intention-to-treat analysis}

The intention-to-treat (ITT) analysis is usually regarded as the preferred approach for the analysis of randomized clinical trials (FDA 1998). The ITT estimator has an interpretation as an estimator of the effect of treatment suggestion on the response variable and, as such, is free of confounding influences. (Note, however, that in order for the ITT effect to have a meaningful interpretation, the instrument should correspond as much as possible to the actual intervention). In the context of our mobile health application, and assuming, once again, stationarity of response data, constance of the causal effect over time, and the core IV assumptions, we have that an unbiased estimator for the effect of treatment assignment on the response is given by Neyman's average causal effect estimator,
\begin{align}
\hat{\beta}_{ITT} &= \frac{\sum_{t=1}^{n} Y_t \, \ind\{ Z_t = 1 \}}{\sum_{t=1}^{n} \ind\{ Z_t = 1 \}} - \frac{\sum_{t=1}^{n} Y_t \, \ind\{ Z_t = 0 \}}{\sum_{t=1}^{n} \ind\{ Z_t = 0\}} \nonumber \\
&= \dfrac{n^{-1} \sum_{t=1}^{n} Z_{t} Y_{t} - (n^{-1} \sum_{t=1}^{n} Z_{t}) (n^{-1} \sum_{t=1}^{n} Y_{t})}{n^{-1} \sum_{t=1}^{n} Z_{t}^2 - (n^{-1} \sum_{t=1}^{n} Z_{t})^2} \nonumber \\
&= \frac{\widehat{\cov}(Z_t, Y_t)}{\widehat{\mbox{Var}}(Z_t)}~,
\label{eq:ittestimator}
\end{align}
which also corresponds to a simple ordinary least squares estimator.

Even though $\hat{\beta}_{ITT}$ estimates the effect of the treatment assignment on the response, as opposed to the effect of the actual treatment received by the participant on the response, it is well known that the ITT comparison still provides a valid statistical test for the null hypothesis of no causal effect of the actual treatment on the response (Rosenberger and Lachin 2002, Hernan and Hernandez-Diaz 2012), as long as, the same core conditions required by the instrumental variable approach hold. (Note that while $Z_t \ci \bfU_t$ holds by construction, and $Z_t \nci X_t$ is expected to hold, the exclusion restriction, $Z_t \ci Y_t \mid \{X_t, \bfU_t\}$, isn't guaranteed to hold, as discussed before.)

\begin{figure}[!h]
\begin{center}
\includegraphics[angle=270, scale = 0.63, clip]{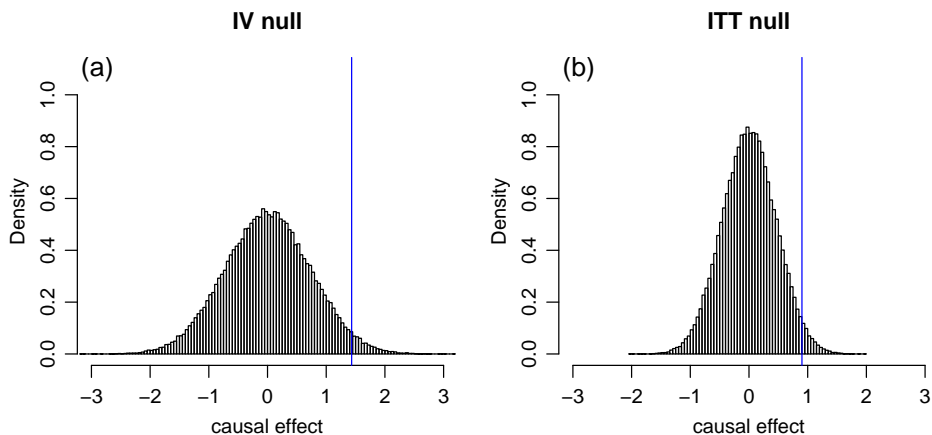}
\caption{Randomization null distributions based on the IV and ITT estimators, computed using the same random permutations of the response data. In spite of the wider spread of the IV null distribution, when compared to the ITT null, in both distributions we have that exactly 5,013 out of the 100,000 permutations of the response data, lead to statistics equal or larger than the observed values in the original data ($\hat{\beta}_{IV} = 1.433$ and $\hat{\beta}_{ITT} = 0.902$, shown by the blue vertical lines). Hence, the p-value derived from both randomization tests are identical and equal to 0.05013.}
\label{fig:randnull}
\end{center}
\end{figure}

As a matter of fact, the randomization tests based on the IV and ITT estimators produce exactly the same p-value if we use the same permutations of the response data in the construction of the randomization null distribution of both tests. Both estimators, share the same numerator, $\widehat{\cov}(Z_t, Y_t)$, which is a function of $Y_t$ and vary with each distinct permutation of the response data employed in the generation of the randomization null distribution, whereas the denominators of the IV and ITT estimators are different (i.e., $\widehat{\cov}(Z_t, Y_t)$ for the IV estimator, and $\widehat{\mbox{Var}}(Z_t)$ for the ITT estimator), but in both cases do not depend on the response data. Therefore, it follows that $\hat{\beta}_{IV} = K^{-1} \, \hat{\beta}_{ITT}$ where, $K = \widehat{\mbox{ACE}}(Z_t \rightarrow X_t) = \widehat{\cov}(Z_t, X_t)/\widehat{Var}(Z_t)$, is not a function of $Y_t$ and is constant across all permutations of the response data used in the construction of the randomization test null distribution. Figure \ref{fig:randnull} shows an illustrative example were we employed the same permutations of the response data in the construction of the IV and ITT null distributions.

We point out, however, that even though adoption of the IV or ITT estimators leads to exactly the same randomization test p-values, the estimates produced by both approaches are different. It is well known that if the treatment has a non-null effect on the response (i.e., $\beta \not= 0$), the ITT approach underestimates the treatment effect when participants do not fully adhere to their assigned treatment, that is, the assigned treatment effect will be closer to zero than the actual treatment effect due to contamination of treatment groups caused by non-compliance. This phenomenon is known as the ``bias towards the null" in placebo-controlled double-blind randomized clinical trials (Hernan and Hernandez-Diaz 2012). The upside of this phenomenon is that, when the treatment has no effect on the response (i.e., $\beta = 0$), this ``bias towards zero" works in favor the ITT approach, which can then correctly estimate the null effect of the treatment on the response.

In the context of our mobile health application, where by construction the treatment assignment mechanism corresponds to a Bernoulli trial with probability of success equal to $P(Z_t = 1) = p = 0.5$, we have that treatment effects estimated by the ITT approach tend to be closer to zero than the estimates from the IV approach since: (i)  if the sample size is not too small, the denominator of $\hat{\beta}_{ITT}$, $\widehat{Var}(Z_t)$, will generally approximate the maximum theoretical variance $\mbox{Var}(Z_t) = 0.25$ (recall that the variance of a Bernoulli random variable with probability of success, $p$, is given by $p \, (1 - p)$, and reaches the maximum of 0.25 when $p = 0.5$); and (ii) the denominator of $\hat{\beta}_{IV}$ will generally be smaller or equal than the denominator of $\hat{\beta}_{ITT}$, $\widehat{\cov}(Z_t, X_t) \leq \widehat{Var}(Z_t)$, since $\widehat{\cov}(Z_t, X_t)$ increases with the amount of compliance, reaching its maximum when the compliance is perfect, in which case, $X_t = Z_t$ for all $t$, and $\widehat{\cov}(Z_t, X_t) = \widehat{Var}(Z_t) \approx 0.25$. This ``bias towards zero" phenomenon is illustrated in Figure \ref{fig:randnull} by the smaller spread around zero of the randomization null distribution based on the ITT estimator, in comparison to the IV randomization null. Supplementary Figure S9 reports a bias comparison between the IV and ITT approaches using the data from the simulation study, and further illustrates this point.

\section{Confidence intervals from randomization tests}

In this section we describe how to build confidence intervals for the causal effect, $\beta$, using the p-values from randomization tests (Garthwaite 1996, Ernst 2004). The procedure is straightforward but requires a considerable amount of computation (which, nonetheless, can be easily parallelized), as we need to test the null, $H_0: \beta = \beta_j$, for each $\beta_j$ on a grid of causal effect values, $\beta_1, \ldots, \beta_J$, and then construct an interval estimate for $\beta$ by considering all $\beta_j$ for which the randomization tests did not reject the null.

\begin{figure}[!h]
\begin{center}
\includegraphics[angle=270, scale = 0.63, clip]{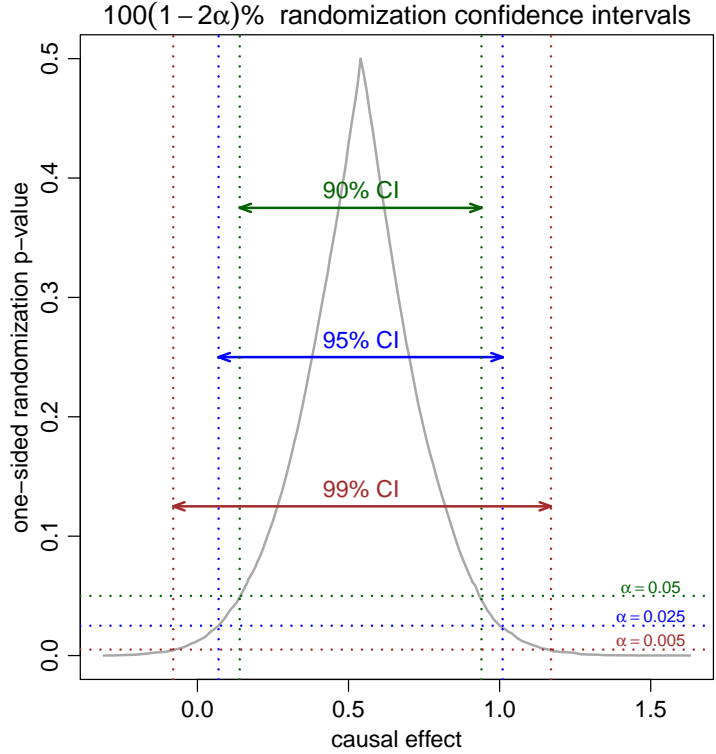}
\caption{Randomization confidence intervals for a data set simulated from an autoregressive model. The estimated effect was $\hat{\beta}_{IV} \approx 0.54$ (true causal effect was 0.5). The grey curve represents the one-sided randomization p-value profile. The x-axis represents the causal effects ($\beta_j$) and the y-axis represents the corresponding randomization p-value for testing the null $H_0: \beta = \beta_j$ against $H_1: \beta > \beta_j$ (for $\beta_j < \hat{\beta}_{IV}$) and $H_1: \beta < \beta_j$ (for $\beta_j > \hat{\beta}_{IV}$). The $100 (1 - 2 \alpha)$\% CI for any given $\alpha$ corresponds to interval inside the p-value profile (along the x-axis) at an $\alpha$ height (along the y-axis). In this example, the 99\% CI (brown) contains 0, illustrating that we do not reject the null $H_0: \beta = 0$ at a $\alpha = 0.01$ (in this example the two tailed p-value was 0.0248). The null $H_0: \beta = 0$ is, nonetheless, rejected at $\alpha$ equal to 0.05 and 0.1, since the respective 95\% (blue) and 90\% (green) confidence intervals do not contain 0.}
\label{fig:randinterval}
\end{center}
\end{figure}

Explicitly, assume for a moment that randomization tests for testing $H_0: \beta = \beta_j$ against one-sided alternative hypothesis $H_1: \beta < \beta_j$ and $H_1: \beta > \beta_j$ are available. Exploring the correspondence between confidence intervals and hypothesis tests, we obtain a $100 (1 - 2 \alpha)$\% confidence interval (CI) for $\beta$ by searching for a lower bound value, $\beta_L$, such that $H_0: \beta = \beta_L$ is rejected in favor of $H_1: \beta > \beta_L$ at a significance $\alpha$, and by searching for an upper bound value, $\beta_U$, such that $H_0: \beta = \beta_U$ is rejected in favor of $H_1: \beta < \beta_U$ at the same significance level (Garthwaite 1996). While an efficient search procedure for finding CI bounds has been proposed in the literature (Garthwaite 1996), the approach requires the specification of the significant level before hand. In order to avoid this constraint, we generate a ``one-sided randomization p-value profile" (Figure \ref{fig:randinterval}) which can be used to determine the $100 (1 - 2 \alpha)$\% CI for any desired $\alpha$ level. This p-value profile is generated as follows: ($i$) compute the observed causal effect estimate, $\hat{\beta}_{IV}$; ($ii$) for each $\beta_j < \hat{\beta}_{IV}$, in a grid of decreasing $\beta_j$ values, compute the randomization p-value from the one-sided test $H_0: \beta = \beta_j$ vs $H_1: \beta > \beta_j$; ($iii$) repeat step $ii$ until a p-value equal to zero is reached; ($iv$) for each $\beta_j > \hat{\beta}_{IV}$, in a grid of increasing $\beta_j$ values, compute the p-value from the one-sided test $H_0: \beta = \beta_j$ vs $H_1: \beta < \beta_j$; ($v$) repeat step $iv$ until a randomization p-value equal to zero is found.

By exploring the equivalence between randomization tests based on the $\hat{\beta}_{IV}$ and $\hat{\beta}_{ITT}$ statistics (described in the previous section), we can easily compute the required randomization p-values using the $\hat{\beta}_{ITT}$ statistic. In other words, instead of directly generating a randomization distribution under the null hypothesis that the causal effect is equal to $\beta_j$ (i.e., $H_0: \beta = \beta_j$), we generate a randomization distribution under the equivalent null hypothesis that the intention-to-treat effect is equal to $\beta_j \, K$ (i.e., $H_0: \mbox{ITT} = \beta_j \, K$), where $K = \widehat{\cov}(Z_t, X_t)/\widehat{\mbox{Var}}(Z_t)$ is constant across all permutations of the response data used in the construction of the randomization null, and connects the $\hat{\beta}_{IV}$ and $\hat{\beta}_{ITT}$ statistics according to the relation $\hat{\beta}_{ITT} = K \, \hat{\beta}_{IV}$. The practical advantage of the test based on ITT effects is that it amounts to a simple two sample location problem for testing whether the difference in average response between the assigned treatment and assigned control groups is equal to $\beta_j \, K$. The implementation of randomization tests for this two sample location problem is straightforward (Garthwaite 1996): we only need to add $\beta_j \, K$ for each $Y_t$ data point in the assigned control group (i.e., $t$ for which $Z_t = 0$), while leaving the response data from the assigned treatment group ($Z_t = 1$) unchanged, and then run a randomization test for testing the null hypothesis that the ITT effect (in this modified version of the data) is equal to zero, against the alternative one-sided hypothesis that the ITT effect is positive (for $\beta_j < \hat{\beta}_{IV}$), and against the alternative that the ITT effect is negative (for $\beta_j > \hat{\beta}_{IV}$).

The grey curve in Figure \ref{fig:randinterval} shows the one-sided p-value profile computed according to the algorithm described above. The x-axis reports the causal effect values (i.e., the $\beta_j$ values) and the y-axis presents the corresponding randomization p-value for testing the null $H_0: \beta = \beta_j$ against $H_1: \beta > \beta_j$ (for $\beta_j < \hat{\beta}_{IV}$) and $H_1: \beta < \beta_j$ (for $\beta_j > \hat{\beta}_{IV}$). The $100 (1 - 2 \alpha)$\% CI of any given $\alpha$ corresponds to interval inside the p-value profile (along the x-axis) at an $\alpha$ height (along the y-axis). The 90\%, 95\%, and 99\% confidence intervals are shown in green, blue, and brown, respectively.

\section{Discussion}

In this paper we proposed the use of instrumental variables in randomized trials with imperfect compliance for causal inference of medication response in mobile health. The present work was motivated by a personalized medicine problem arising from the mPower study, and represents an improvement over a previous contribution (Chaibub Neto \textit{et al} 2016b), which was based on the naive assumption of no unobserved confounders.

A practical objective of the present work was to evaluate the empirical power of the randomization test, and assess the feasibility of the IV approach in the context of mobile health. We were particularly interested in evaluating the empirical power under varying amounts of compliance by the study participants. Our simulations suggest that, at least for the reasonably wide range of parameter values evaluated in this work, the IV approach is indeed well powered, even when the degree of compliance is moderate. Additionally, in practice, it seems reasonable to expect moderate to high compliance levels since the simple electronic suggestion that the participant should perform the activity task either before or after taking medication does not seem to cause much of a burden to a participant, and we expect that the participants will be able to comply with the suggested treatment most of the time.

As discussed in Section 5, the randomization test based on the IV estimator is equivalent to the randomization test based on the ITT estimator. Therefore, a simple ITT analysis could be used in place of the IV approach. We point out, however, that often times a researcher will be interested in estimating the strength of the causal effect when it turns out that a statistical test suggests the effect is different from zero. In this situation, and if data checks suggest that the core IV conditions and additional parametric assumptions required by the IV approach seem to hold (for instance we can empirically check whether the instrument and treatment variables are associated, whether the association between the treatment and response variables seem to be approximately linear, whether the response and treatment data time series seem to be stationary, and etc), then the IV estimator might be preferred over the ITT estimator, as the latter tends to be biased towards zero (Supplementary Figure S9a). Otherwise, both estimators might be biased.

The present work relies on the mechanism-based account of causation championed by Pearl (2000). Application of Pearl's interventional calculus in the context of time series models was first proposed by Eichler and Didelez (2007, 2010) and Eichler (2012). We point out, however, that their approach is not based in IVs, and their goal was to model the effect of an intervention in one component of a multivariate time series model on another component at a later point in time.

Randomized clinical trials have been used in the evaluation of the effectiveness of mobile interventions for health behavior change or disease management in several areas including: smoking cessation, physical activity/diet, sexual health, alcoholism, CPR interventions, medication adherence, diabetes management, asthma and chronic obstructive pulmonary interventions, hypertension, and psychological interventions (Free \textit{et al} 2013). These trials were, nonetheless, tailored to the conventional framework of population medicine, in opposition to the personalized medicine focus of the present work.

Single case research designs (Franklin \textit{et al} 1997) have also been used in the context mobile health (Dallery \textit{et al} 2013). These studies generally include a small number of participants subjected to periods of treatment intercalated with periods of non-treatment, and the longitudinal data of each participant is usually analyzed separately. The goal, nonetheless, is to establish the efficacy of a treatment or intervention in a given cohort, and not at the personalized level.

The recently proposed micro-randomization designs (Liao \textit{et al} 2015, Dempsey \textit{et al} 2015) generalize single-case designs by allowing more traditional statistical analysis of multiple participants concomitantly, under a population medicine framework. Micro-randomization designs adopt a potential outcomes framework and allow the inference of proximal time dependent causal effects of just-in-time mobile interventions. A goal of just-in-time interventions is to promote behavior change, and help participants manage stressful situations in the moment the intervention is needed (e.g., a participant of a drinking cessation program might benefit from a motivational message popping up on the smartphone screen when in close proximity to a liquor store). Hence, micro-randomization trials address a different problem from the one motivating the present work.

Although the approach proposed in this paper represents a step towards causal inference for medication response in mobile health applications, an important pragmatic challenge still remains. Because participants are not blinded to the treatment they are actually receiving, it is not really possible the tell whether an observed ``medication response" is truly caused by a medication effect, or because the participant tends to perform better after taking medication due to psychosomatic effects, or due to a combination of both medication and psychosomatic effects. Hence, the unambiguous determination of the medication effect is still contingent on the assumption that the study participant is not prone to psychosomatic effects. We point out, however, that Chaibub Neto (2016) recently proposed an instrumental variable approach to disentangle treatment and placebo effects in unblinded trials, which could be potentially employed in our motivating application.

Even though the personalized medicine problem motivating the present work involves the self-selection of study participants, most of the perils and pitfalls involved in web-based epidemiological studies and surveys (Keiding and Louis 2016) are avoided by our focus on the estimation of participant-specific treatment effects. Finally, it is important to point out that while the proposed IV approach is well equipped to handle confounding, it is still vulnerable to selection bias.

\section{Acknowledgements}

ECN, BMB, MK, SHF, ADT, LO, and LM acknowledge funding from the Robert Wood Johnson Foundation. RLP was supported by NCI grant P01 CA53996.

\beginsupplement


\section{Supplementary figures}

$ $

\begin{figure}[!h]
\begin{center}
\includegraphics[angle=270, scale = 0.55, clip]{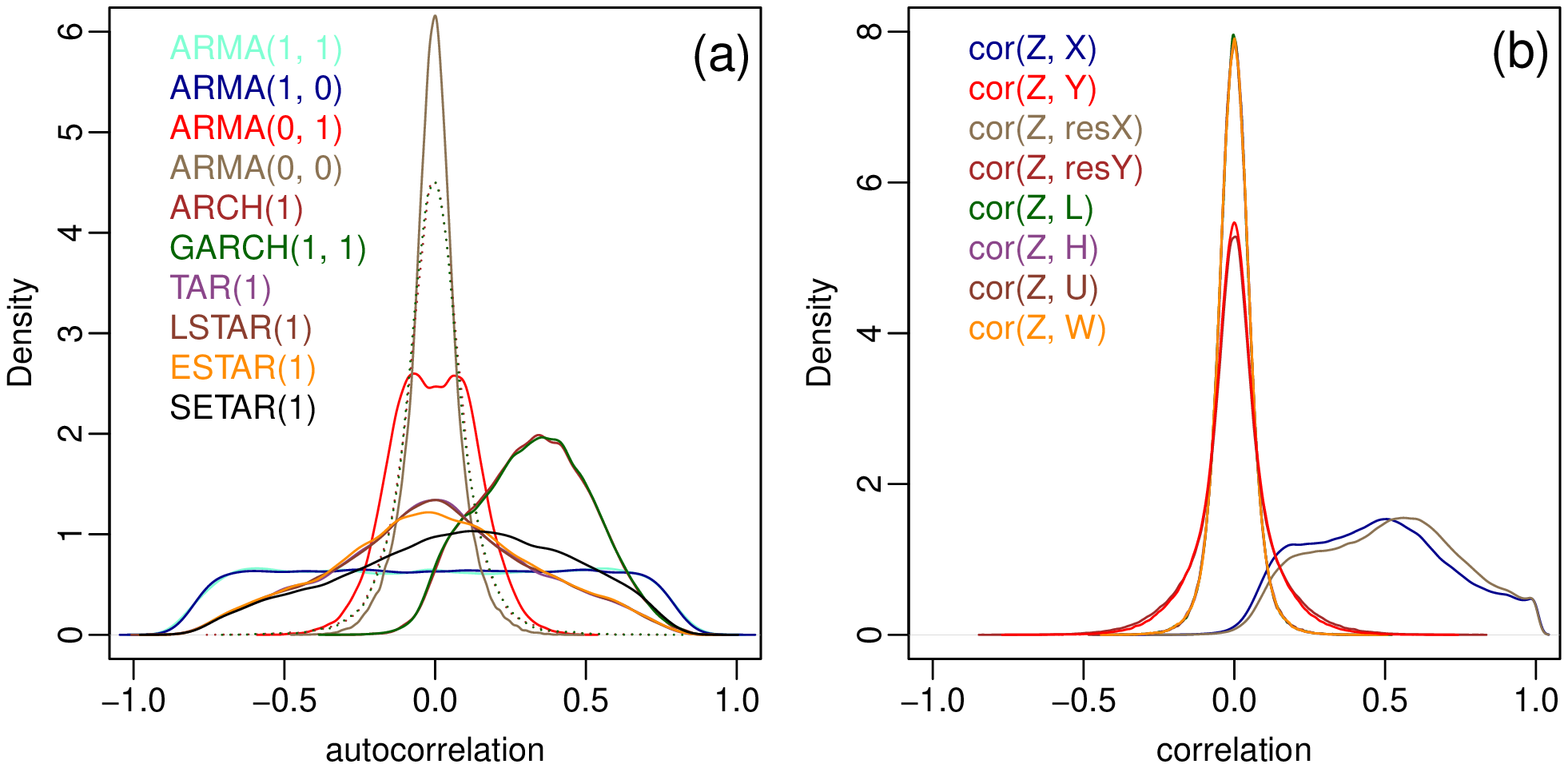}
\caption{Panel a presents the distribution of the (lag = 1) autocorrelations for the response variables for all 10 models in Table \ref{tab:tsmodels}. For the ARCH(1) and GARCH(1, 1) models the dotted and full line curves show, respectively, the lag 1 autocorrelations for the response and squared response measurements. We see that except for the ARMA(0, 0) model (which does not impose a serial correlation structure over the response variable) all other models generate autocorrelated responses. Panel b presents the distributions of the correlations between the instrumental variable, $Z$, and all other variables. The densities were estimated using all 800,000 simulated data sets. The densities clearly show mostly positive correlations between the instrumental and treatment variables (blue and silver curves), moderate correlations between the instrumental and response variables (red and brown curves), and weak correlations between the instrument and all other measured covariates and unmeasured confounders.}
\label{supplefig:acfycorz}
\end{center}
\end{figure}

\begin{figure}[!h]
\begin{center}
\includegraphics[angle=270, scale = 0.55, clip]{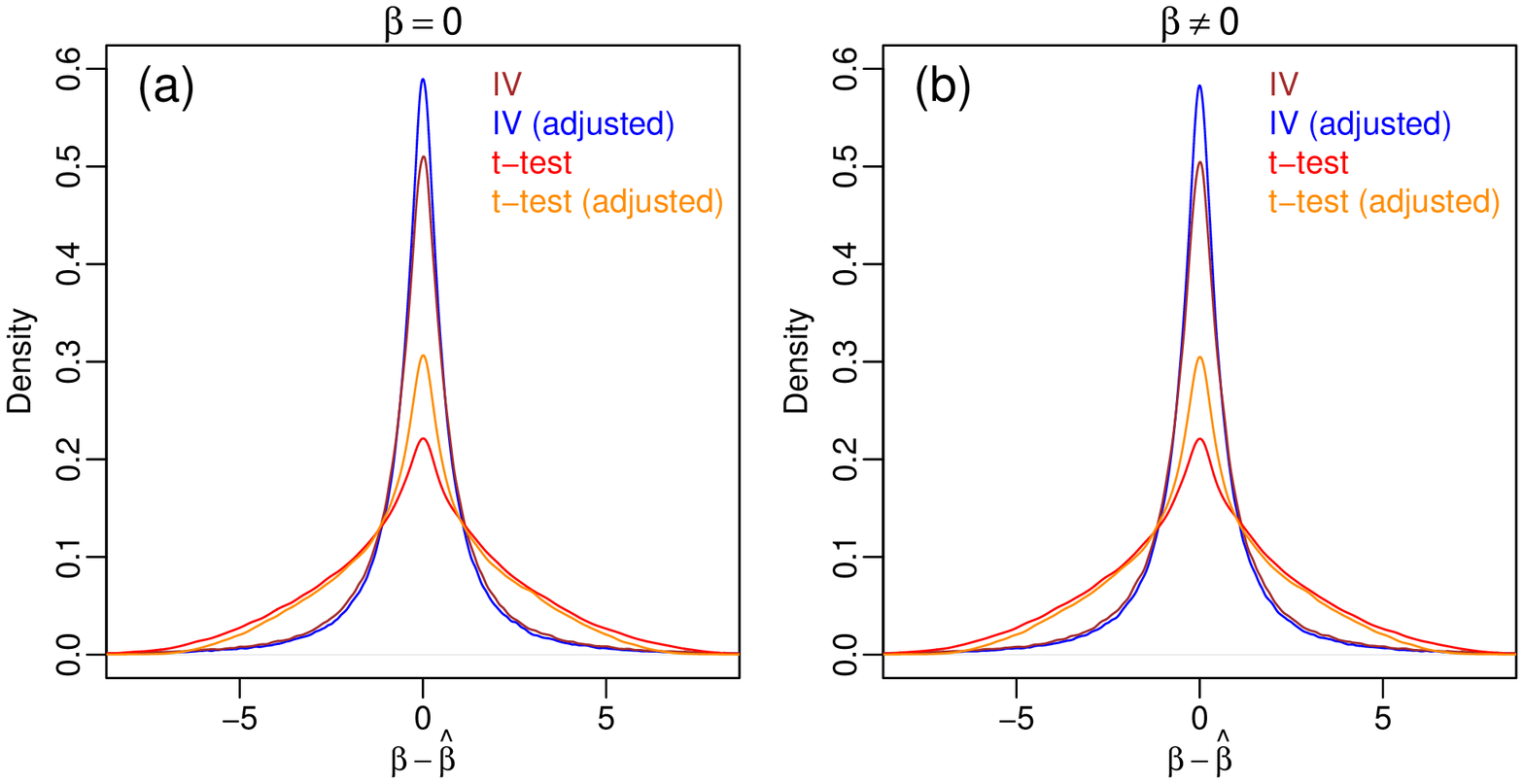}
\caption{Comparison of the bias of the t-tests and IV estimators. Panels a and b show the results for data simulated under the null and alternative hypothesis, respectively. Note that the $\hat{\beta}$ estimates generated by the t-tests tend to show larger bias than the estimates generated by the instrumental variable approach, as illustrated by the heavier trails of the t-test (red) and adjusted t-test (orange) distributions, when compared to the IV approaches.}
\label{supplefig:deltabeta}
\end{center}
\end{figure}

\begin{figure}[!h]
\begin{center}
\includegraphics[angle=270, scale = 0.6, clip]{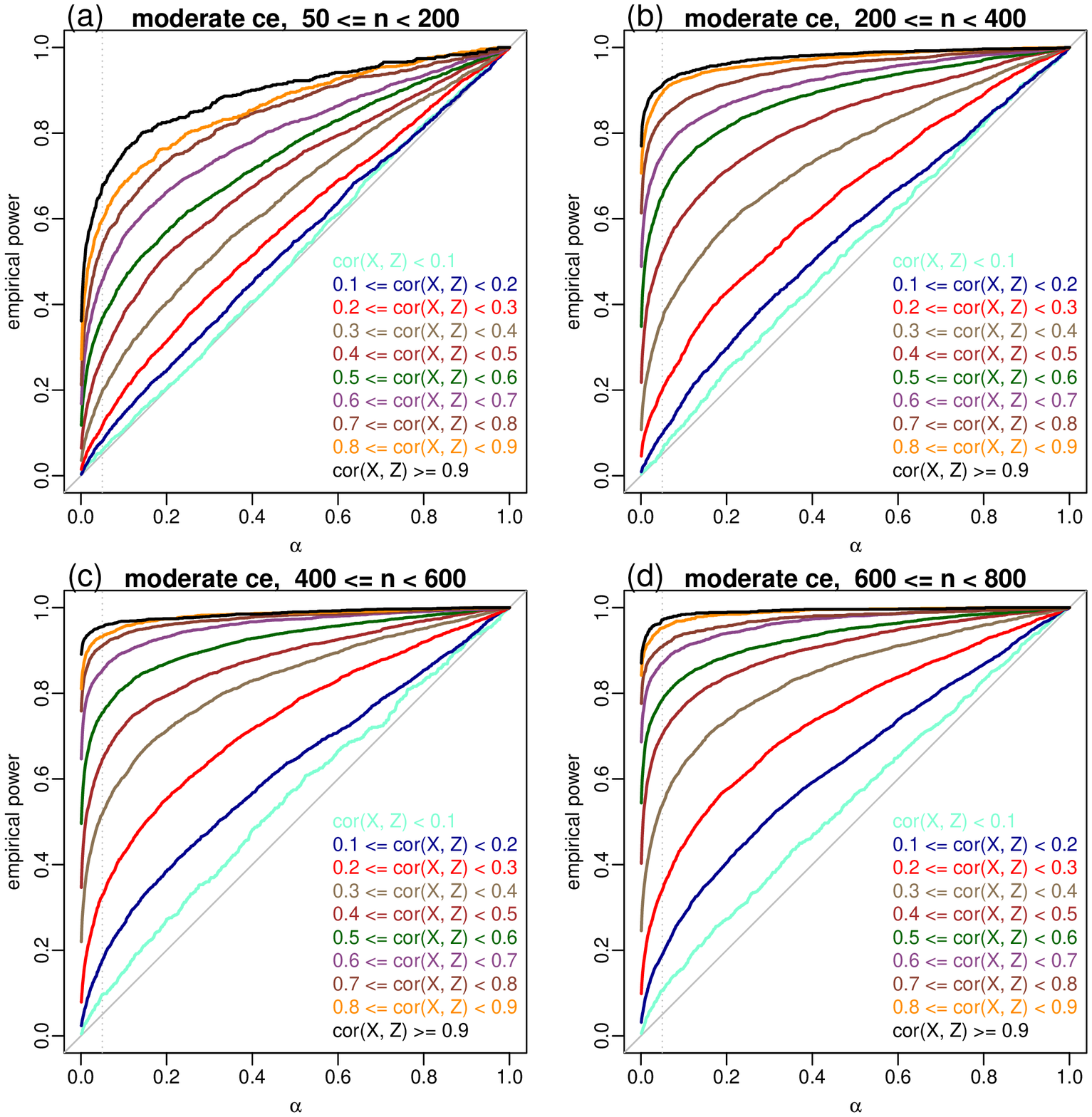}
\caption{Empirical power curves for the adjusted IV causal effect estimator stratified according to cor(X, Z) and to the sample size, n, for simulated data sets generated with moderate causal effect, $1 \leq |\beta| < 3$. The vertical dotted line is set at $\alpha = 0.05$.}
\label{supplefig:powerbycorxzmoderate}
\end{center}
\end{figure}

\begin{figure}[!h]
\begin{center}
\includegraphics[angle=270, scale = 0.6, clip]{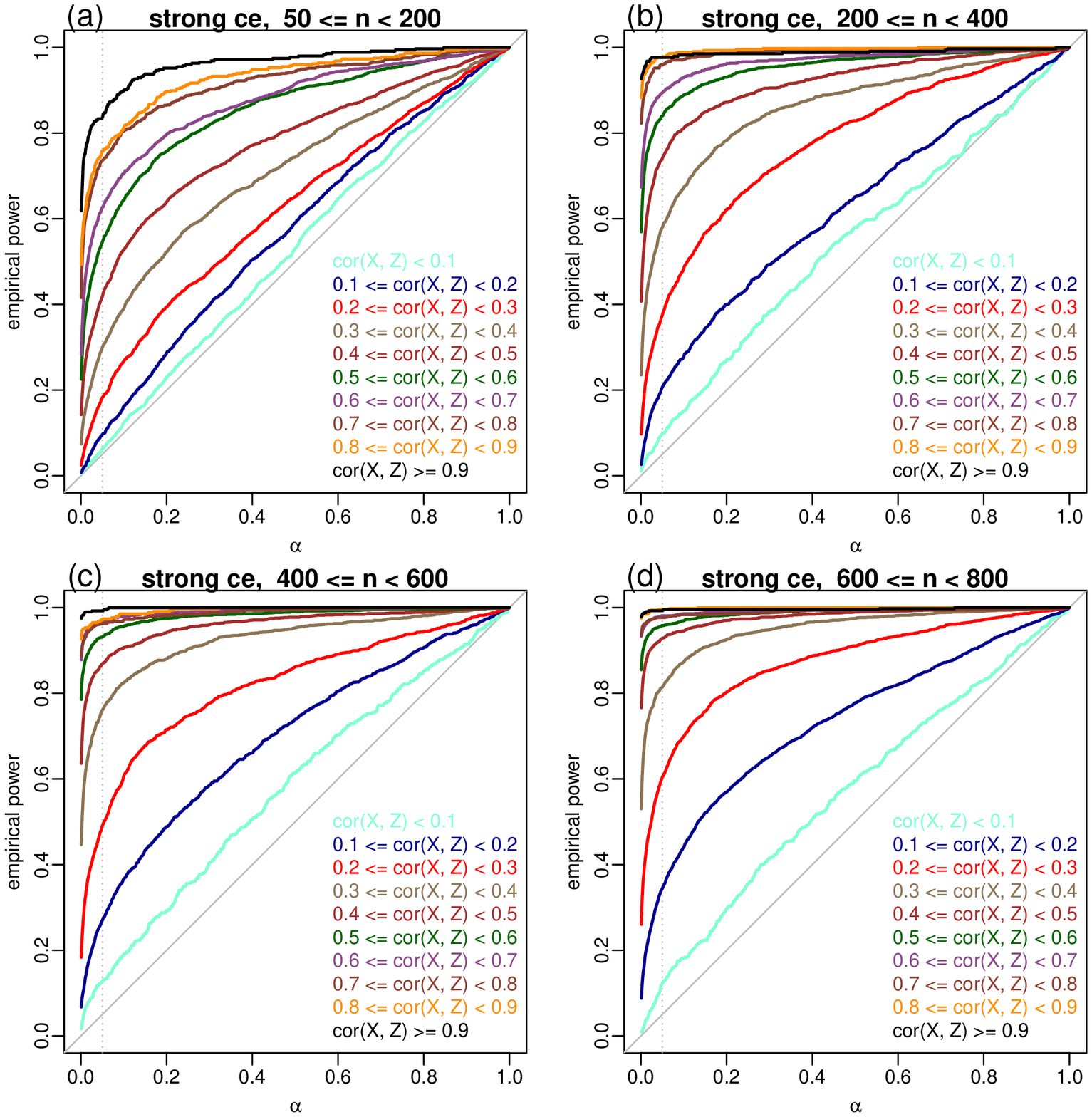}
\caption{Empirical power curves for the adjusted IV causal effect estimator stratified according to cor(X, Z) and to the sample size, n, for simulated data sets generated with strong causal effect, $|\beta| \geq 3$. The vertical dotted line is set at $\alpha = 0.05$.}
\label{supplefig:powerbycorxzstrong}
\end{center}
\end{figure}

\begin{figure}[!h]
\begin{center}
\includegraphics[angle=270, scale = 0.5, clip]{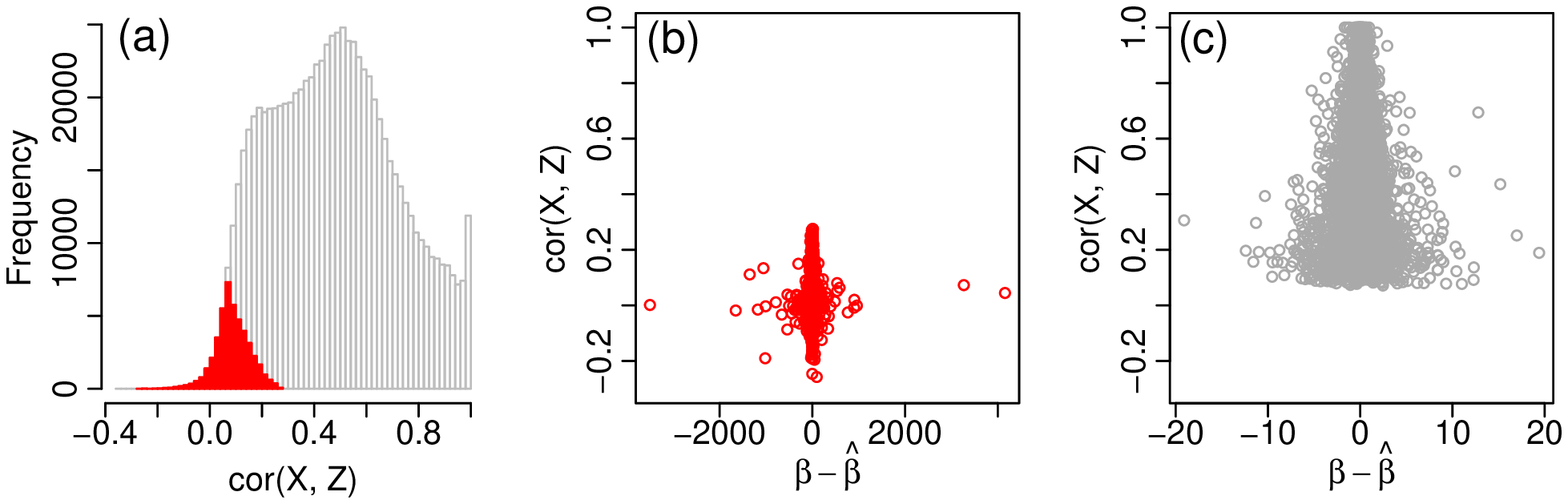}
\caption{Panel a shows the distribution of $\mbox{cor}(Z, X)$ across all simulated data sets. Highlighted in red is the distribution of $\mbox{cor}(Z, X)$ for the data sets for which the correlation is not statistically different from zero at a significance level equal to 0.05. Panels b and c show scatter-plots of $\beta - \hat{\beta}$ against $\mbox{cor}(Z, X)$ for 5,000 randomly selected data sets, for which the correlation is not statistically different from zero (panel b), and for which the correlation is statistically different from zero (panel c). It is clear that the IV estimator can generate highly biased estimates when the $Z_t \nci X_t$ assumption is violated (note that the x-axis range in panel b is orders of magnitude larger than in panel c).}
\label{supplefig:filteredcor}
\end{center}
\end{figure}

\begin{figure}[!h]
\begin{center}
\includegraphics[angle=270, scale = 0.45, clip]{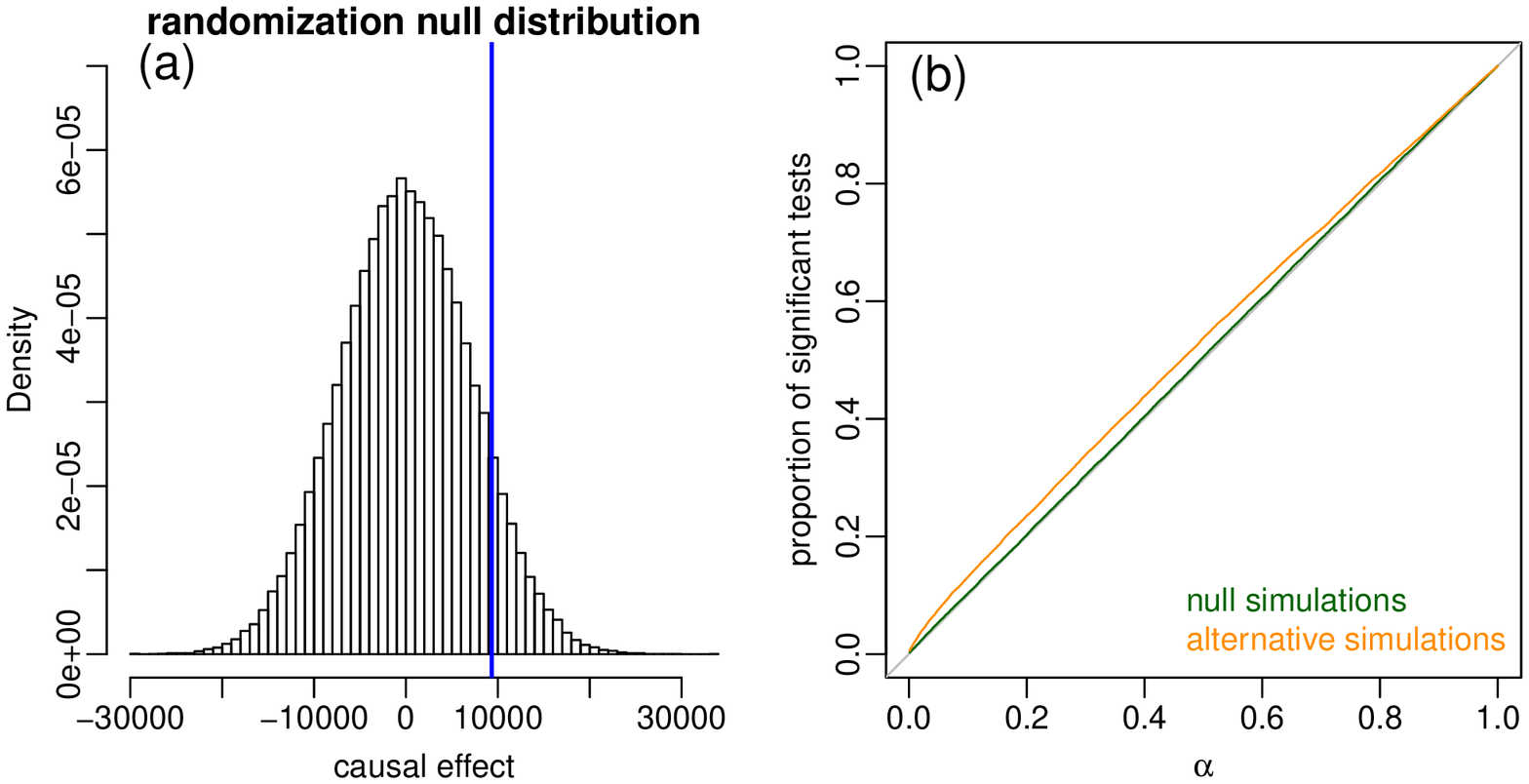}
\caption{Empirical type I error rates and empirical power, when compliance is very low (i.e., $\cov(X, Z)$ is close to zero). Panel a shows the randomization null distribution for one simulated data set where the causal effect estimated by the adjusted IV approach was extremely high ($\hat{\beta} = 9329.01$). Even though the estimate is super inflated by the small $\widehat{\cov}(X, Z) \approx -0.002$ estimate in the denominator of $\hat{\beta}$, the randomization test is still non-significant. This protection follows from the fact that we only shuffle the response data in the generation of the randomization test null distribution, but keep the association of the instrumental and treatment variables intact, so that the denominator of the IV estimator is always the same in all shufflyings of the data used to generate the null. Panel b shows the empirical type I error rates (dark green) and empirical power (dark orange) for the simulations for which $cor(X,Z)$ was not statistically different from zero (encompassing 22,902 data sets simulated under the null and 22,923 under the alternative hypothesis). Note that the type I error rate is still well controlled but the test lacks power to detect causal effects when they exist. The practical consequence of these observations is that it is still safe to apply the randomization tests for very low levels of compliance if the goal is simply to help a physician flag patients which respond to medication (as low compliance will drastically reduce the statistical power but won't lead to spurious findings).}
\label{supplefig:behavior}
\end{center}
\end{figure}

\begin{figure}[!h]
\begin{center}
\includegraphics[angle=270, scale = 0.51, clip]{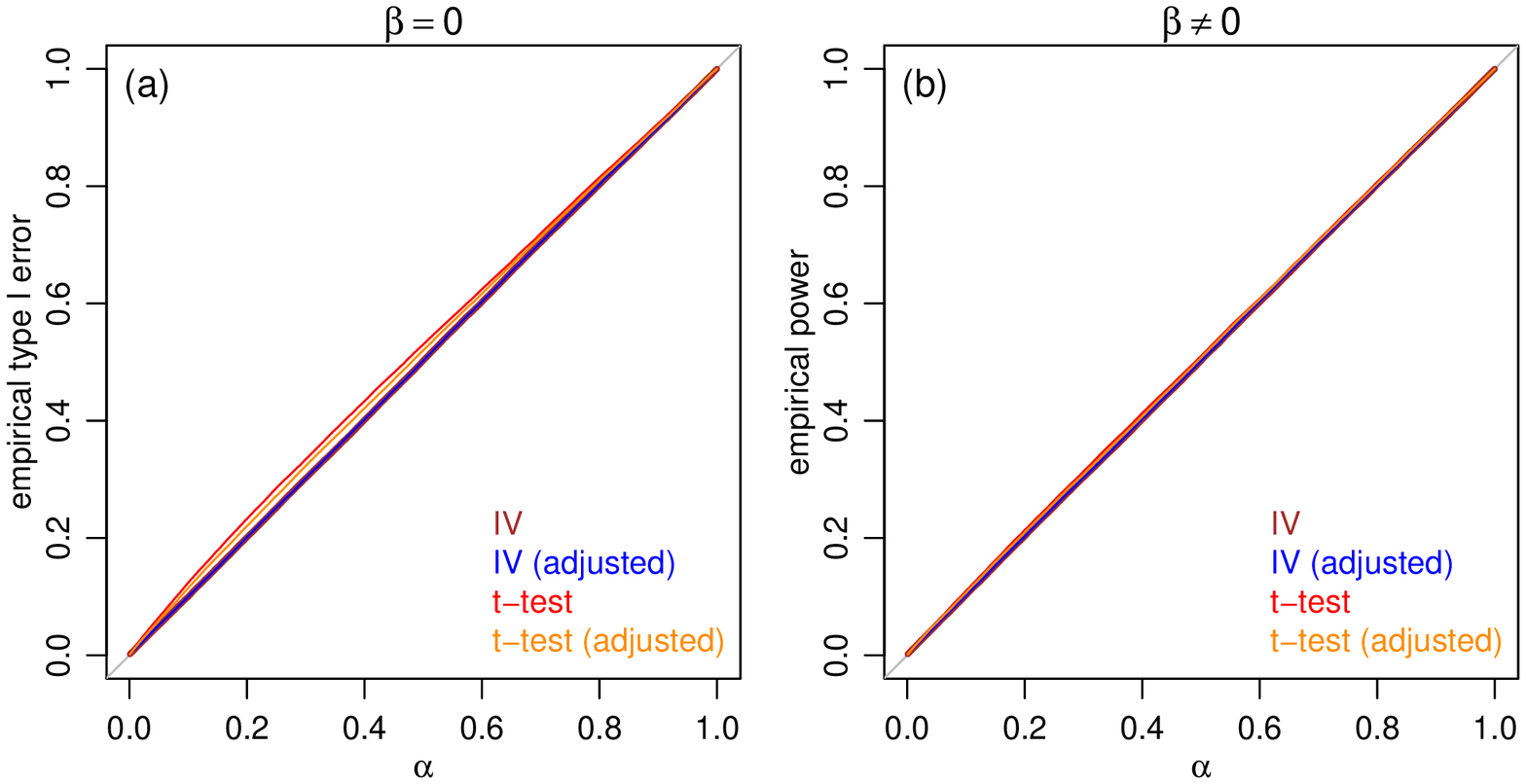}
\caption{Empirical type I error rates and empirical power, under violations of the stationarity assumption. Panel a shows that the type I error rates for the straight and adjusted IV approaches still match the nominal levels. Panel b, on the other hand, shows that the empirical power to detect the causal effect is completely wiped out. For these simulations we only considered the ARMA(1, 1), ARMA(1, 0), TAR(1), and SETAR(1) models. We set the autoregressive parameters, $\phi_1$, $\phi_{1,1}$, $\phi_{1,2}$, and $\rho$, to 1 in order to generate non-stationary data while still avoiding explosive processes. For all remaining parameters, we adopted the same parameter ranges of Table 3. As before, we generated the data under the eight simulation settings of Table 2.}
\label{supplefig:nonstat}
\end{center}
\end{figure}

\begin{figure}[!h]
\begin{center}
\includegraphics[angle=270, scale = 0.51, clip]{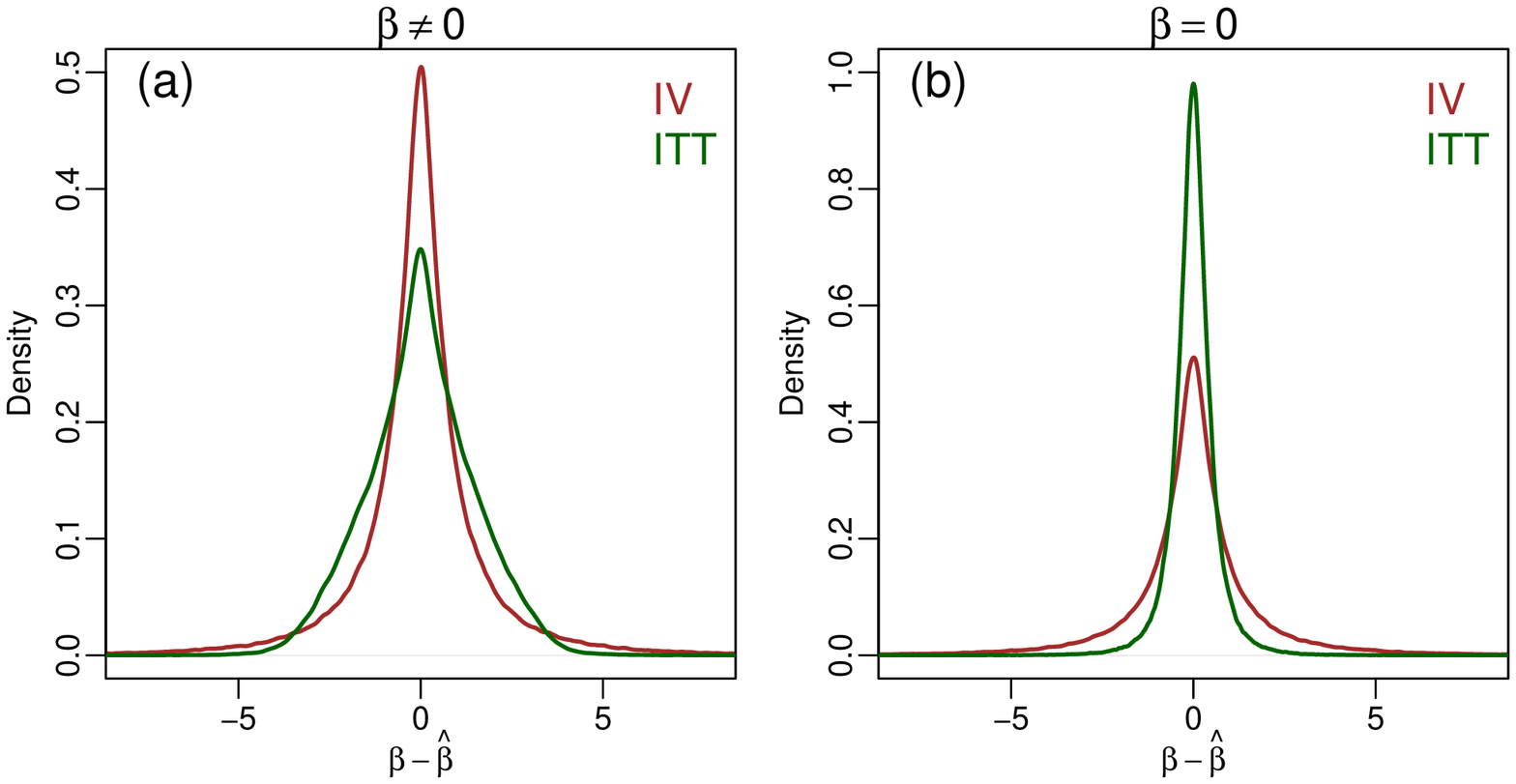}
\caption{Bias comparison between the IV and ITT approaches. Panel a compares the IV and ITT approaches using data generated under the alternative hypothesis, $H_1 : \beta \not= 0$. Results show that the ITT approach tends to generate more biased estimates of $\beta$ than the IV approach, whose density puts more probability mass at values close to zero (brown density). Panel b shows the results for data generated under the null hypothesis, $H_0 : \beta = 0$, where the ITT approach tends to generate less biased estimates of $\beta = 0$ than the IV approach. Note that while the IV densities (brown curves) are similar in both panels (with peak close to 0.5 at $\beta - \hat{\beta} \approx 0$), the variation in the shape of the ITT densities (dark-green curves) is due to the ``bias towards zero" phenomenon, which works against the ITT estimator under $H_1 : \beta \not= 0$ (panel a), but on its favor under $H_0 : \beta = 0$ (panel b).}
\label{supplefig:ivitt}
\end{center}
\end{figure}

\clearpage
\section{Appendix A: minimizing bias due to the time of the day that the activity is performed}

In the context of our motivating application, it is the participant who decides the time of the day that he/she will perform the activity task. Because of this particularity, it is possible that the suggested treatment, $Z_t$, will influence both the treatment actually adopted by the participant, $X_t$, and the the time of the day that the participant performs the task (for now on, denoted by TOD). Explicitly, suppose that a participant takes medication regularly at about the same time every day (say, around noon). If the randomized suggestion asks the participant to perform the task before taking medication (i.e., $Z_t = 0$), and the participant decides to comply with $Z_t$ (i.e., $X_t = 0$), the participant will perform the task in the morning. Similarly, if the participant complies with a suggestion to perform the task after taking medication, then the participant will perform the task in the afternoon. Hence, if the participant tends to comply with the suggested treatment, we have that most tasks performed before medication will be done in the morning, and most tasks performed after medication will be done in the afternoon. Therefore, it is not really possible to determine whether the medication has an effect on the outcome (task performance), or if an association between $X_t$ and $Y_t$ is actually caused by daily physiological variations due to circadian rhythms (e.g., the participant might inherently tend to do better in the afternoon than in the morning), or caused by daily routine activities for which the TOD might be a surrogate measurement. For instance, features extracted from accelerometer data recorded during a walking task (where the participant is asked to put its phone in the front pocket and walk in a straight line for a fixed period of time) are affected by the clothing the participant is using during the activity. Hence, if the participant tends to perform the ``before medication" walking activity in the morning, while wearing comfortable pajamas and sandals, and the ``after medication" activity later in the day, while wearing tight paints and shoes, it is not possible to tell if the association between $X_t$ and $Y_t$ is due to the medication or to the clothing the participant is using.

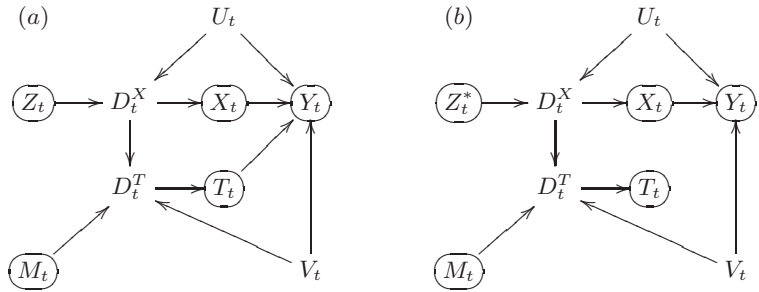
\begin{figure}[!h]
$$
{\footnotesize \xymatrix@-0.6pc{
(a) & & U_t \ar[dl] \ar[dr] &&&  (b) & & U_t \ar[dl] \ar[dr]  \\
*+[F-:<10pt>]{Z_t} \ar[r] & D^{X}_t \ar[r] \ar[d] & *+[F-:<10pt>]{X_t} \ar[r] & *+[F-:<10pt>]{Y_t} &&  *+[F-:<10pt>]{Z_t^\ast} \ar[r] & D^{X}_t \ar[r] \ar[d] & *+[F-:<10pt>]{X_t} \ar[r] & *+[F-:<10pt>]{Y_t}  \\
& D^{T}_t \ar[r] & *+[F-:<10pt>]{T_t} \ar[ur] & &&   & D^{T}_t \ar[r] & *+[F-:<10pt>]{T_t} & &   \\
*+[F-:<10pt>]{M_t} \ar[ru] & & & V_t \ar[uu] \ar[llu] &&  *+[F-:<10pt>]{M_t} \ar[ru] & & & V_t \ar[uu] \ar[llu]  \\
}}
$$
\caption{Panel a shows the causal DAG representing the influence of the treatment suggestions, $Z_t$, on the treatment adopted, $X_t$, on time of the day that the activity was performed (TOD), $T_t$, and on the outcome, $Y_t$. Please, refer to the text for a description of the variables and explanations. Panel b shows the case where the modified suggestions, $Z_t^\ast$, generate a narrow time window between the before and after medication activity tasks, so that the TOD is unlikely to influence the outcome.}
\label{fig:toddag}
\end{figure}

Graphically, the causal DAG representing the influence of the TOD is shown is Supplementary Figure \ref{fig:toddag}a, where $M_t$ represents the usual time that the participant takes its medication and $T_t$ represents the TOD. The variable $D^X_t$ represents the participant's decision about the treatment he/she will actually take (note that, in addition to the suggested treatment, $Z_t$, the unobserved confounder of the treatment/outcome relation, $U_t$, also influences the participant's decision). The arrow from $T_t$ to $Y_t$ indicates the influence of circadian rhythms or daily routine activities (for which the TOD is a surrogate variable) on the performance on the activity task. The variable $D^T_t$ represents the participant's decision about what time he/she will perform the activity task. Note that $D^T_t$ is a function of $D^X_t$ and $M_t$ since if a participant takes medication around the same time every day, then the time the participant decides to performs the activity task will depend on the treatment that the participant decided to take. The model also allows for the presence of unmeasured confounders, $V_t$, influencing $D^T_t$ and the outcome. The arrows from $D^X_t$ to $X_t$ and from $D^T_t$ to $T_t$ indicate that the participant's decisions precedes the actual actions and that unexpected events can change the actual treatment and the time the activity is performed.

It is clear from Supplementary Figure \ref{fig:toddag}a that even though the TOD is not technically a confounder of the medication effect (in the sense that a confounder of the treatment/outcome relation is usually defined as a variable that is a common cause of $X_t$ and $Y_t$) failing to adjust for the influence of the TOD in the outcome, leads to a biased estimate of the medication effect because the core instrumental variable assumption \textit{iii}, $Z_t \ci Y_t \mid \{X_t, U_t\}$, is violated due to the additional path $Z_t \rightarrow D^X_t \rightarrow D^T_t \rightarrow T_t \rightarrow Y_t$ mediated by the TOD.

\begin{figure}[!h]
\includegraphics[angle=270, scale = 0.6, clip]{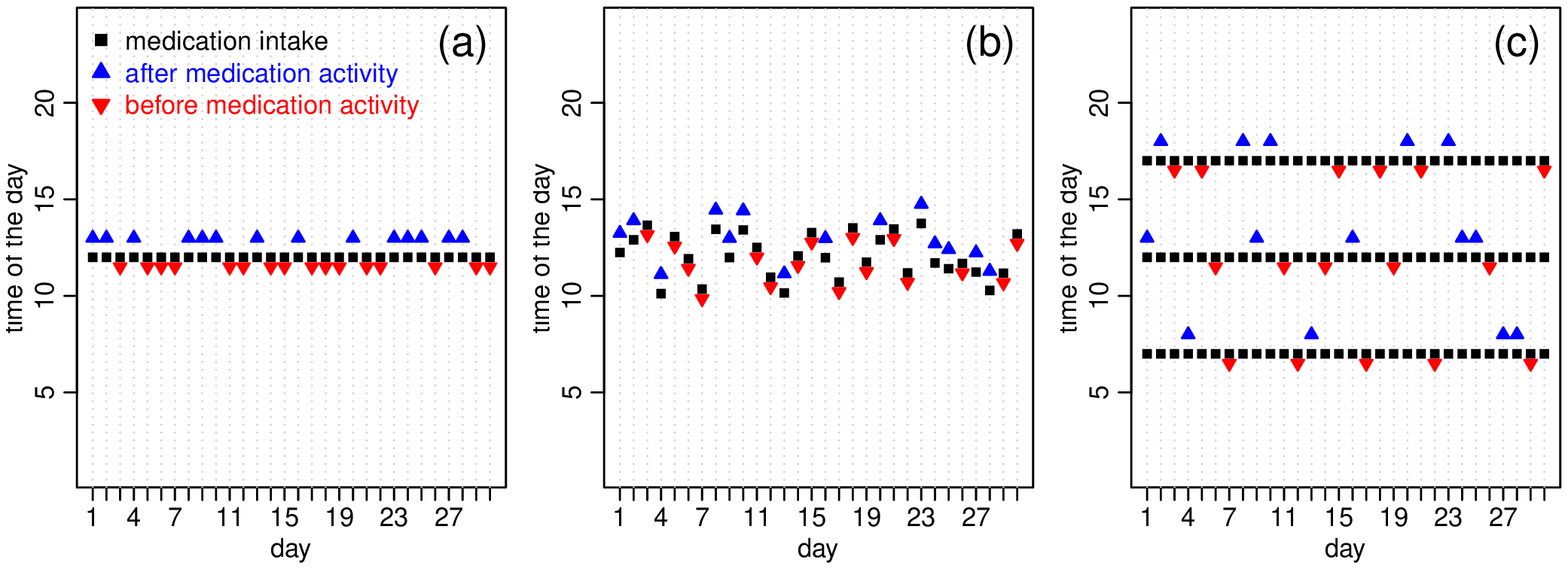}
\caption{Influence of the time the participant usually takes its medication on the randomization of the treatments across the time of the day. In all three panels the modified treatment suggestions (``please, perform the activity task right before taking your medication" and ``please, perform the activity task about one hour after taking your medication") are randomized across the days (and the participant fully complies with the suggestions). Panel a shows an example where the participant takes medication once per day at noon. All the tasks performed before medication are done a little earlier than noon, while all tasks performed after medication are done at 1pm. Because the time window between the before and after medication activity tasks is very narrow, it is unlikely that physiological variations during the day or routine habits will bias the effect of the medication. Panel b shows an example where the participant takes the medication once per day between 10am and 2pm. In this case the participant is even less prone to confounding effects due to TOD, since the ``before medication" task in one day can be performed later than the ``after medication" task of another day, and vice-versa (note how the red and blue triangles are intermingled). Panel c illustrates the case where the participant takes the medication, regularly, but at 7am, noon, and 5pm. As long as the suggestions are also randomized across these three times of the day, bias due to the TOD will be less of an issue as well.}
\label{fig:randpatterns}
\end{figure}

In applications based on fast acting drugs (such was L-dopa used in treating Parkinson's disease, which usually produces an effect is less than an hour), a simple solution to decrease the potential influence of the TOD is to slightly modify the $Z_t$ suggestions. Instead of simply asking the participant to perform the activity task either before or after taking medication, we could ask the participants to perform the ``before medication" task just before taking the medication, and the ``after medication" task at a fixed period of time after taking the medication. For instance, if the drug is known to reach its peak effect in about one hour after intake, then the ``after medication" suggestion should be ``please, perform the activity task about one hour after taking your medication". In this way, we have that for the participants that take their medication at the same time everyday, the difference in the time that the participant performs the before and after medication tasks will be about one hour only, so that the influence of physiological variations during the day will be minimized. Supplementary Figure \ref{fig:toddag}b illustrates this case, where the narrow time window makes it reasonable to assume that, at least approximately, there is no influence of $T_t$ on $Y_t$. (Of course, however, if the participant usually takes medication early in the morning and routinely performs the ``before medication" task using pajamas and sandals, and the ``after medication" task one hour after taking medication, and after having changed clothes, the medication effect might still be biased by the ``clothing effect". However, by keeping a narrow time window between the before and after medication activity tasks we decrease the likelihood of coming across biased effects due to routine habits.)

Observe that data from participants that take their medication according to a more irregular schedule are less prone to confounding effects due to TOD, since the ``before medication" task in one day can be performed later than the ``after medication" task of another day, and vice-versa. Similarly, participants that take the medication, regularly, but at multiple times per day, are also less prone to confounding due to TOD, as long as, the suggestions are also randomized across the medication intake events (e.g., ``please, perform the task just before taking medication, at the second time of the day that you take your medication" or ``please, perform the task one hour after taking medication, at the first time of the day that you take your medication"). Supplementary Figure \ref{fig:randpatterns} shows a few cartoon examples illustrating these cases.

\section{Appendix B: non-parametric identification of the causal effects of $Z_t$ on $X_t$ and of $Z_t$ on $Y_t$}

Let $\mG$ represent a dynamic DAG for which the IV assumptions \textit{i} to \textit{iii} described in the main text hold, but otherwise arbitrary. Note that, in this case, $Z_t$ will always be an exogenous variable in $\mG$ (i.e., $Z_t$ has no parents in $\mG$). Let $\bfV$ represent the set of all variables in $\mG$, and $\bfA = \bfV \setminus \{Y_t, Z_t\}$. Observe that the set $\bfA$ includes instrumental and response variables over all time points other than $t$, treatment and time specific confounders and covariates over all time points, as well as, ubiquitous confounders and covariates.

Since $Z_t$ is an exogenous variable in $\mG$, we can factor the joint distribution of $\bfV$ as,
\begin{equation}
P(y_t, \, \bfa, \, z_t) = P(y_t, \, \bfa \mid z_t) \, P(z_t)~.
\end{equation}
Although the conditional joint distribution, $P(y_t, \, \bfa \mid z_t)$, can be further factorized according to $\mG$, we don't need to specify the factorization explicitly when determining the post-intervention distribution for the intervention $do(Z_t = z')$, since application of the truncated factorization formula reduces to removing $P(z_t)$, and replacing $z_t$ by $z'$ in the remaining conditional distributions, so that,
\begin{equation}
P\big(y_t, \, \bfa \mid do(Z_t = z')\big) = P(y_t, \, \bfa \mid z')~,
\end{equation}
independent of how $P(y_t, \, \bfa \mid z')$ can be further factorized. The marginal post-intervention distribution is given by,
\begin{equation}
P\big(y_t \mid do(Z_t = z')\big) = \sum_{\bfa} P(y_t, \, \bfa \mid z') = P(y_t \mid z')~,
\label{eq:marginalcausaleffect}
\end{equation}
where the summation over $\bfa$ is simply a notation for all the summations or integrations over each one of the variables in the set $\bfA$.

The average causal effect of $Z_t$ on $Y_t$ is then given by,
\begin{align}
\mbox{ACE}(Z_t \rightarrow Y_t) &= E\big(Y_t \mid do(Z_t = 1)\big) - E\big(Y_t \mid do(Z_t = 0)\big) \\ \nonumber
&= E\big(Y_t \mid Z_t = 1\big) - E\big(Y_t \mid Z_t = 0\big),
\end{align}
where the second equality follows from (\ref{eq:marginalcausaleffect}). Assuming stationarity of the $Y_t$ time series we have that a large sample non-parametric estimate of the expectation $E\big(Y_t \mid Z_t = z'\big)$ is given by,
\begin{equation}
\dfrac{\sum_{t=1}^{n} Y_t \, \ind\{ Z_t = z' \}}{\sum_{t=1}^{n} \ind\{ Z_t = z' \}}~,
\end{equation}
so that,
\begin{align}
\widehat{\mbox{ACE}}(Z_t \rightarrow Y_t) &= \frac{\sum_{t=1}^{n} Y_t \, \ind\{ Z_t = 1 \}}{\sum_{t=1}^{n} \ind\{ Z_t = 1 \}} - \frac{\sum_{t=1}^{n} Y_t \, \ind\{ Z_t = 0 \}}{\sum_{t=1}^{n} \ind\{ Z_t = 0\}} \nonumber \\
&= \frac{\sum_{t=1}^{n} Y_t \, Z_t}{\sum_{t=1}^{n} Z_t} - \frac{\sum_{t=1}^{n} Y_t \, (1 - Z_t)}{\sum_{t=1}^{n} (1 - Z_t)} \nonumber \\
&= \frac{n^{-1} \sum_{t=1}^{n} Z_{t} Y_{t} - (n^{-1} \sum_{t=1}^{n} Z_{t}) (n^{-1} \sum_{t=1}^{n} Y_{t})}{(n^{-1} \sum_{t=1}^{n} Z_{t}) (1 - n^{-1} \sum_{t=1}^{n} Z_{t})}~.
\end{align}

Now, let $\bfB = \bfV \setminus \{X_t, Z_t\}$. Then, by a similar rational it follows that,
\begin{equation}
P\big(x_t \mid do(Z_t = z')\big) = \sum_{\bfb} P(x_t, \, \bfb \mid z') =  P(x_t \mid z')~,
\end{equation}
and
\begin{equation}
\widehat{\mbox{ACE}}(Z_t \rightarrow X_t) = \frac{n^{-1} \sum_{t=1}^{n} Z_{t} X_{t} - (n^{-1} \sum_{t=1}^{n} Z_{t}) (n^{-1} \sum_{t=1}^{n} X_{t})}{(n^{-1} \sum_{t=1}^{n} Z_{t}) (1 - n^{-1} \sum_{t=1}^{n} Z_{t})}~.
\end{equation}

\section{Appendix C: selection bias in the context of mobile health}

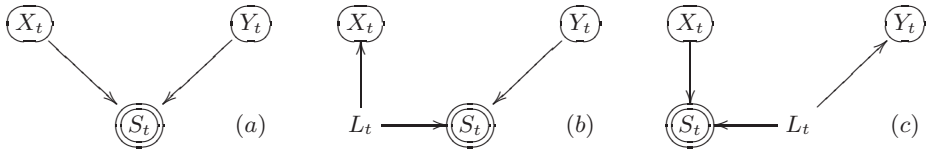
\begin{figure}[!h]
$$
\xymatrix{
*+[F-:<10pt>]{X_t} \ar[dr] && *+[F-:<10pt>]{Y_t} \ar[dl] & *+[F-:<10pt>]{X_t} && *+[F-:<10pt>]{Y_t} \ar[ld] & *+[F-:<10pt>]{X_t} \ar[d] && *+[F-:<10pt>]{Y_t} \\
& *+<9pt>[F=:<10pt>]{S_t} & (a) & L_t \ar[r] \ar[u] & *+<9pt>[F=:<10pt>]{S_t} & (b) & *+<9pt>[F=:<10pt>]{S_t} & L_t \ar[l] \ar[ru] & (c) \\
}
$$
\caption{Causal structures giving rise to selection bias. As pointed by Hernan, Hernandez-Diaz, and Robins (2004), the term selection bias represents several distinct biases, but the causal structure underlying all these sort of biases is essentially the same, namely, selection bias arises when we condition on a collider node, $S_t$, which is a common effect of two variables: one of which is either the exposure or a cause of the exposure, and the other is either the outcome or a cause of the outcome. Panels a, b, and c show three examples where conditioning in the collider node, $S_t$, leads to selection bias. Note that conditioning on $S_t$ leads to spurious association between $X_t$ and $Y_t$ because it allows information to flow between the $X_t$ and $Y_t$ by opening up paths between these variables. For instance, conditioning on $S_t$ opens up the path $X_t \rightarrow S_t \leftarrow Y_t$ in panel a, the path $X_t \leftarrow L_t \rightarrow S_t \leftarrow Y_t$ in panel b, and the path $X_t \rightarrow S_t \leftarrow L_t \rightarrow Y_t$ in panel c. The double framing around $S_t$ indicates that the analysis is conditional on $S_t = 1$.}
\label{fig:selectionbias1}
\end{figure}

In addition to confounding, selection bias is another major obstacle to the validity of causal inference in clinical studies. Selection bias is hard to detect in observational and experimental settings, and cannot be removed by randomization (Bareinboim, Tian, and Pearl 2014). It occurs when samples are preferentially selected to the data set, and is a consequence of spurious association between the exposure and outcome variable generated by conditioning on a collider node (Hernan, Hernandez-Diaz, and Robins 2004; Bareinboim, Tian, and Pearl 2014). Supplementary Figure \ref{fig:selectionbias1} shows three causal DAGs where $X_t$ is not a cause of $Y_t$, but where conditioning on the collider node $S_t$ generates a spurious association between the $X_t$ and $Y_t$ variables. The variable $S_t$ represents a selection variable, where $S_t = 1$ indicates inclusion in the data set, and $S_t = 0$ represents exclusion (i.e., the data is missing). Note that by restricting the analysis to the data recorded in the data set, we are effectively conditioning the analysis on $S_t = 1$, what induces a spurious association between $X_t$ and $Y_t$. Panel a represents the case where both exposure and the outcome influence whether or not the data will be included in the data set. Panel b shows the case where a latent variable, $L_t$, influences the exposure, while both outcome and $L_t$ influence the inclusion in the data set. Panel c represents the case where $L_t$ influences the outcome, while both exposure and $L_t$ influence the selection to the data set.

In the context of our motivating example, the $X_t$ variable can easily influence the selection of the sample into the data set. For instance, suppose that a participant feels very debilitated before taking medication. In this case, it is more likely that the participant will be unable to complete the activity task when unmedicated than when medicated. The $Y_t$ variable, on the other hand, does not seem to be able to influence $S_t$. Please recall that in our application the outcome $Y_t$ is simply a quantitative feature extracted from the activity task (e.g., number of taps in fixed time interval) and measures the participant's performance in the task. (It is very different in nature than the outcome of, for example, a case-control study, where the outcome represents an indicator of disease status, and the arrow from the outcome to the selection variable indicates that cases are more likely to be selected to the study than non-cases.) Observe, nonetheless, that selection bias can still play a role in our application in situations where the outcome and selection variables are both influenced by a common latent variable, as illustrated in Supplementary Figure \ref{fig:selectionbias1}c.

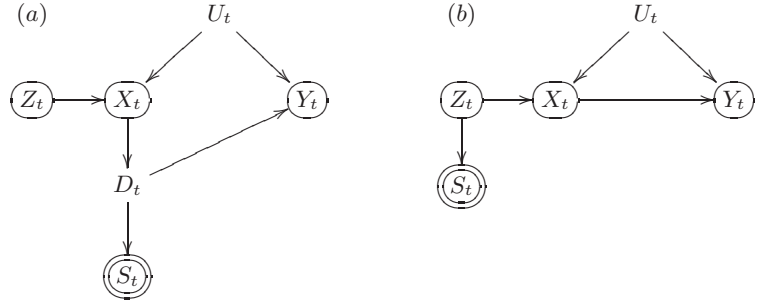
\begin{figure}[!h]
$$
{\footnotesize \xymatrix@-0.5pc{
(a) && U_t \ar[dl] \ar[dr] &&& (b) && U_t \ar[dl] \ar[dr] && \\
*+[F-:<10pt>]{Z_t} \ar[r]  & *+[F-:<10pt>]{X_t} \ar[d] && *+[F-:<10pt>]{Y_t} && *+[F-:<10pt>]{Z_t} \ar[r] \ar[d] & *+[F-:<10pt>]{X_t} \ar[rr] && *+[F-:<10pt>]{Y_t} \\
 & D_t \ar[rru] \ar[d] & && & *+<9pt>[F=:<10pt>]{S_t} &  && \\
 & *+<9pt>[F=:<10pt>]{S_t} & & & \\
}}
$$
\caption{Examples, where informative missing data/selection mechanisms do not lead to selection bias. Because $S_t$ is not a collider, in both panels, conditioning on $S_t$ does not lead to selection bias in these examples.}
\label{fig:selectionbias2}
\end{figure}

We point out, however, that informative missing data mechanisms do not necessarily lead to selection bias. Supplementary Figure \ref{fig:selectionbias2} shows two plausible examples that might play a role in our motivating application. Panel a revisits the case where a participant feels very debilitated before taking medication, and is more likely to produce missing data when $X_t = 0$, than when $X_t = 1$. In this example, the latent variable $D_t$ corresponds to the debilitation state of the participant, and shows that the treatment influences the debilitation state which, by its turn, influences the likelihood that the participant will be able to perform/complete the task and contribute a data point (i.e., $X_t \rightarrow D_t \rightarrow S_t$). Furthermore, it is also likely that the debilitation status influences the performance of the participant in the activity task (i.e., $D_t \rightarrow Y_t$). Observe, however, that contrary to the DAGs in Supplementary Figure \ref{fig:selectionbias1}, conditioning on $S_t$ does not generate further spurious association between $X_t$ and $Y_t$, since $S_t$ is not a collider (or a descendent of a collider) in the DAG in Supplementary Figure \ref{fig:selectionbias2}a. Hence, selection bias will not be an issue in this example, even though data collection is subjected to an informative missing data mechanism.

Supplementary Figure \ref{fig:selectionbias2}b shows still another example where an informative missing data mechanism does not lead to selection bias. Now, suppose a participant has a very competitive personality and enjoys winning and excelling in any activity that he/she participates. If the participant feels he/she achieves better performance on the activity task when medicated ($X_t = 1$), the participant might be more prone to skip the activity in the days he/she is assigned to do it before taking medication $(Z_t = 0)$, than when assigned to do it after medication ($Z_t = 1$), generating, in this way, another non-random missing data pattern. In this example, the selection variable is directly influenced by the treatment suggestion (i.e., $Z_t \rightarrow S_t$), and, once again, $S_t$ is not a collider.

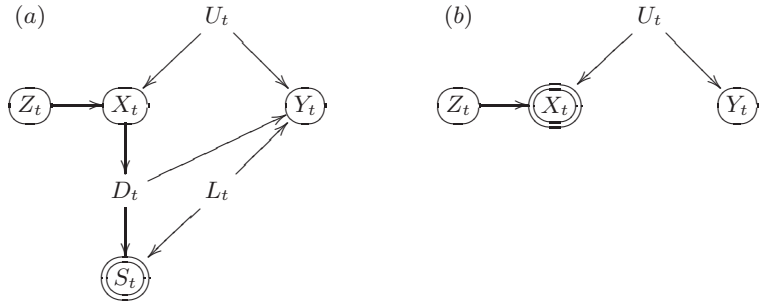
\begin{figure}[!h]
$$
{\footnotesize \xymatrix@-0.5pc{
(a) && U_t \ar[dl] \ar[dr] &&& (b) && U_t \ar[dl] \ar[dr] & \\
*+[F-:<10pt>]{Z_t} \ar[r]  & *+[F-:<10pt>]{X_t} \ar[d] && *+[F-:<10pt>]{Y_t} && *+[F-:<10pt>]{Z_t} \ar[r]  & *+<9pt>[F=:<10pt>]{X_t} && *+[F-:<10pt>]{Y_t} \\
 & D_t \ar[rru] \ar[d] & L_t \ar[ld] \ar[ru] & &&& &  \\
 & *+<9pt>[F=:<10pt>]{S_t} & &&   \\
}}
$$
\caption{Examples of selection bias. Panel a shows an example where an informative missing data/selection mechanism leads to selection bias. Because $S_t$ is a collider, conditioning on $S_t$ opens an additional path, $X_t \rightarrow D_t \rightarrow S_t \leftarrow L_t \rightarrow Y_t$, between $X_t$ and $Y_t$, biasing the causal effect of $X_t$ on $Y_t$. Panel b shows an example of selection bias due to selecting on the treatment. Note that while this type of bias is not an issue in our motivating application (since our treatment variable is genuinely binary), selecting on the treatment can be another important source of bias in other mobile health applications. For example, suppose that the goal is to compare two distinct medications, but where the participant sometimes doesn't take any of the two drugs. In this case the treatment variable has three levels, namely, ``drug A", ``drug B", and ``no drug". Suppose, as well, that the drugs are equally effective, so that $\beta = E[Y_t \mid X_t = A] - E[Y_t \mid X_t = B] = 0$, and we don't see an effect of $X_t$ on the outcome $Y_t$. By restricting the analysis to days where the participant took either drug A or B, we are effectively conditioning on the collider node $X_t$, and generating an spurious association between $Z_t$ and $Y_t$, by opening the path $Z_t \rightarrow X_t \leftarrow U_t \rightarrow Y_t$. Consequently, the IV estimator will produce a biased estimate since its numerator, $\widehat{\cov}(Z_t, Y_t)$, will capture this spurious association.}
\label{fig:selectionbias3}
\end{figure}

Supplementary Figure \ref{fig:selectionbias3}a, on the other hand, shows an example where an informative missing data/selection mechanism indeed leads to selection bias. Here, suppose that the latent variable $L_t$ represents the depression level of the participant on day $t$. It is reasonable to expect that a participant might be more inclined to skip an activity task when depressed than when feeling well (i.e., $L_t \rightarrow S_t$). Additionally, a participant might achieve better performance on an activity task when felling well, as opposed to when feeling depressed (i.e., $L_t \rightarrow Y_t$). In this example, the informative missing data mechanism, due to the joint influence of the debilitation and depression level, will lead to selection bias, since conditioning on the collider $S_t$ leads to spurious additional association between $X_t$ and $Y_t$ by opening the path $X_t \rightarrow D_t \rightarrow S_t \leftarrow L_t \rightarrow Y_t$. This selection mechanism induces a bias in the instrumental variable estimator since its numerator, $\cov(Z_t, Y_t)$, ends up capturing association generated by both paths $Z_t \rightarrow X_t \rightarrow D_t \rightarrow Y_t$ and $Z_t \rightarrow X_t \rightarrow D_t \rightarrow S_t \leftarrow L_t \rightarrow Y_t$, instead of only by the $Z_t \rightarrow X_t \rightarrow D_t \rightarrow Y_t$ path.

Finally, we point out that while selection bias due to selecting on the treatment (Swanson et al. 2015) is not an issue in our motivating application (since our treatment variable is genuinely binary), selecting on the treatment can be another important source of bias in mobile health applications where the analysis is restricted to two levels of a treatment variable that is not actually binary, as illustrated in Supplementary Figure \ref{fig:selectionbias3}b.



\begin{thebibliography}{20}

\bibitem{angrist2001}
Angrist J, Krueger A (2001). Instrumental variables and the search for identification: from supply and demand to natural experiments. \textit{Journal of Economic Perspectives}, \textbf{15}, 69-85.

\bibitem{baiocci2014}
Baiocchi M, Cheng J, Small DS (2014). Instrumental variable methods for causal inference. \textit{Statistics in Medicine}, \textbf{33}, 2297-2340.

\bibitem{beasley1997}
Beasley TM, Allison DB, Gorman BS (1997). The potentially confounding effects of cyclicity: identification, prevention, and control. In Franklin RD, Allison DB, Gorman BS (eds), Design and Analysis of Single-case Research. Lawrence Erlbaun Associates, New Jersey.

\bibitem{bollerslev1986}
Bollerslev T (1986). Generalized Autoregressive Conditional Heteroskedasticity. \textit{Journal of Econometrics}, \textbf{31}, 307-327.

\bibitem{bot2016}
Bot B et al (2016). The mPower study, Parkinson disease mobile data collected using ResearchKit. \textit{Scientific Data}, 3:160011 doi:10.1038/sdata.2016.11

\bibitem{boxjenkins1994}
Box G, Jenkins GM, Reinsel GC (1994). \textit{Time Series Analysis: Forecasting and Control}. Third edition. Prentice-Hall.

\bibitem{bowden1990}
Bowden RJ, Turkington DA (1990). \textit{Instrumental Variables}. Cambridge University Press.

\bibitem{chaibubneto2016a}
Chaibub Neto E (2016a). Using instrumental variables to disentangle treatment and placebo effects in blinded and unblinded randomized clinical trials influenced by unmeasured confounders. \textit{Scientific Reports}, \textbf{6}, 37154, doi:10.1038/srep37154.

\bibitem{chaibubneto2016b}
Chaibub Neto E et al (2016b). Personalized hypothesis tests for detecting medication response in Parkinson disease patients using iPhone Sensor data. \textit{Pacific Symposium on Biocomputing}, \textbf{21}, 273–284.

\bibitem{dallery2013}
Dallery J, Cassidy R, Raiff B (2013). Single-case experimental designs to evaluate novel technology-based health interventions. \textit{Journal of Medical Internet Research}, \textbf{15}, e22.

\bibitem{dempsey2015}
Dempsey W, Liao P, Klasnja P, Nahum-Shani I, Murphy SA (2015). Randomized trials for the Fitbit generation. \textit{Significance}, \textbf{12}, 20-23.

\bibitem{didelez2010}
Didelez V, Meng S, Sheehan NA (2010). Assumptions of IV methods for observational epidemiology. \textit{Statistical Science}, \textbf{25}, 22-40.

\bibitem{engle1982}
\textsc{Engle RF} (1982). Autoregressive conditional heteroscedasticity with estimates of variance of United Kingdom inflation. \textit{Econometrica}, \textbf{50}, 987-1008.

\bibitem{eichler2007}
Eichler M, Didelez V (2007). Causal reasoning in graphical time series models. \textit{Proc. 23rd Conf. on Uncertainty in Artificial Intelligence}, 19–22 July, Vancouver, BC (eds Parr R, van der Gaag L). Corvallis, OR, AUAI Press.

\bibitem{eichler2010}
Eichler M, Didelez V (2010). On Granger-causality and the effect of interventions in time series. \textit{Life Time Data Analysis}, \textbf{16}, 3–32.

\bibitem{eichler2012}
Eichler M (2012). Causal inference in time series analysis. In Causality (eds Berzuini C, Dawid AP, Bernardinelli L), pp. 327–354. Chichester, UK, Wiley

\bibitem{ernst2004}
Ernst MD (2004). Permutation methods: a basis for exact inference. \textit{Statistical Science}, \textbf{19}, 676-685.

\bibitem{fda1998}
Food and Drug Administration (1998). International Conference on Harmonisation; Guidance on Statistical Principles for Clinical Trials. \textit{Federal Register}, 63, 49583–98.

\bibitem{fisher1990}
Fisher LD, Dixon DO, Herson J, Frankowski RK, Hearron MS, Peace KE (1990). Intention-to-treat in clinical trails. In Peace KE, editor. Statistical Issues in Drug Research and Development. New York, Marcel Dekker, 331-350.

\bibitem{franklin1997}
Franklin RD, Allison DB, Gorman BS (1997). \textit{Design and Analysis of Single-Case Research}. Lawrence Erlbaun Associates, New Jersey.

\bibitem{free2013}
Free C, et al (2013). The effectiveness of mobile-health technology-based health behavior change or disease management interventions for health care consumers: a systematic review. \textit{PLoS Medicine}, \textbf{10}, e1001362.

\bibitem{stephen2015}
Friend SH (2015). App-enabled trial participation: tectonic shift or tepid rumble? \textit{Science Translational Medicine} {\bf 7}, 297ed10.

\bibitem{garthwaite1996}
Garthwaite PH (1996). Confidence intervals from randomization tests. \textit{Biometrics}, \textbf{52}, 1387-1393.

\bibitem{greenland2000}
Greenland S (2000). An introduction to instrumental variables for epidemiologists. \textit{International Journal of Epidemiology}, \textbf{29}, 722-729.

\bibitem{herman2006}
Hernan MA, Robins JM (2006). Instruments for causal inference: an epidemiologist's dream? \textit{Epidemiology}, \textbf{17}, 360-372.

\bibitem{hernan2012}
Hernan MA, Hernandez-Diaz S (2012). Beyond the intetion-to-treat in comparative effectiveness research. \textit{Clinical Trials}, \textbf{9}, 48-55.

\bibitem{johnson1990}
Johnson M, Moore L, Ylvisaker D (1990). Minimax and maximin distance designs. \textit{Journal of Statistical Planning and Inference}, \textbf{26}, 131-148.

\bibitem{keiding2016}
Keiding N, Louis TA (2016). Perils and potentials of self-selected entry to epidemiological studies and surveys (with discussion). \textit{Journal of the Royal Statistical Society, Series A}, \textbf{179}, 319-376.

\bibitem{liao2015}
Liao P, Klasnja P, Tewari A, Murphy SA (2015). Micro-randomized trials in mHealth. arXiv:1504.00238v1.

\bibitem{pearl2000}
Pearl J (2000). \textit{Causality: Models, Reasoning, and Inference}. Cambridge University Press.

\bibitem{rosenberger2002}
Rosenberger WF, Lachin JM (2002). \textit{Randomization in Clinical Trials, Theory and Practice}. John Wiley \& Sons, New York.

\bibitem{santner2003}
Santner TJ, Williams BJ, Notz WI (2003). \textit{The Design and Analysis of Computer Experiments}. Springer Verlag, New York.

\bibitem{schork2015}
Schork NJ (2015). Personalized medicine: time for one-person trials. Nature, 520: 609-611.

\bibitem{timeseries2011}
Shumway RH, Stoffer DS (2011). \textit{Time Series Analysis and Its Applications With R Examples}. Third Edition, Springer.

\bibitem{tar1978}
Tong H (1978). On a threshold model. In Chen C (ed.) Pattern Recognition and Signal Processing. Amsterdam, Sijthoff \& Noordhoff, pp. 575-586.

\bibitem{setar1980}
Tong H, Lim KS (1980). Threshold autoregression, limit cycles and cyclical data. \textit{Journal of the Royal Statistical Society, Series B}, \textbf{42}, 245-292.

\bibitem{topol2012}
Topol E (2012). The orientation of medicine today: population versus the individual. In, The Creative Destruction of Medicine. Basic Books, New York.

\bibitem{trister2016}
Trister AD, Dorsey ER, Friend SH (2016). Smartphones as new tools in the management and understanding of Parkinson's disease. \textit{npj Parkinson's Disease}, 16006.

\bibitem{star2002}
Van Dijk D, Terasvirta T, Franses PH (2002). Smooth transition autoregressive models - a survey of recent developments. \textit{Econometric Reviews}, \textbf{21}, 1-47.

\end{thebibliography}

\begin{thebibliography}{20}

\bibitem{bareinboim2014}
\textsc{Bareinboim E, Tian J, Pearl J} (2014). Recovering from selection bias in causal and statistical inference. \textit{Proceedings of the Twenty-Eighth AAAI Conference on Artificial Inteligence}, 2410-2416.

\bibitem{herman2004}
\textsc{Hernan MA, Hernadez-Diaz S, Robins JM} (2004). A structural approach to selection bias. \textit{Epidemiology}, \textbf{15}, 615-625.

\bibitem{swanson2015}
\textsc{Swanson SA, Robins JM, Miller M, Hernan MA} (2015) Selecting on treatment: a pervasive form of bias in instrumental variable analyses. \textit{American Journal of Epidemiology}, \textbf{181}, 191-197.

\end{thebibliography}
\end{document}